\documentclass[11pt]{article}
\usepackage{graphics}
\usepackage{psfrag}
\usepackage{psfig}
\usepackage{float}
\usepackage{latexsym}
\usepackage[german,USenglish]{babel}
\newcommand{\beq}{\begin{equation}}
\newcommand{\eeq}{\end{equation}}
\newcommand{\beqa}{\begin{eqnarray}}
\newcommand{\eeqa}{\end{eqnarray}}
\newcommand{\nn}{\nonumber \\ }
\newcommand{\bma}{\begin{array}{cc}}
\newcommand{\ema}{\end{array}}

\def\3{{\ss}}

\def\vek #1 {\overrightarrow {#1}}
\newcommand{\fet}[1]{\mbox{\boldmath $#1$}}

\begin{document}

\hfill {\tiny FZJ-IKP(TH)-2002-18}

\vspace{0.8cm}

\begin{center}

{{\Large\bf Three--Nucleon Forces from Chiral Effective Field Theory}}

\end{center}

\vspace{.3in}

\begin{center}

{\large 
E. Epelbaum,$^\dagger$\footnote{email: 
                           evgeni.epelbaum@tp2.ruhr-uni-bochum.de}
A. Nogga,$^\star$\footnote{email:
                           anogga@physics.arizona.edu}
W. Gl\"ockle,$^\dagger$\footnote{email:
                           walter.gloeckle@tp2.ruhr-uni-bochum.de}
H. Kamada,$^\ast$\footnote{email:
                           kamada@mns.kyutech.ac.jp}
Ulf-G. Mei{\ss}ner,$^\ddagger$\footnote{email: 
                           u.meissner@fz-juelich.de}
H. Wita{\l}a$^\circ$}\footnote{email:
                           witala@if.uj.edu.pl}

\bigskip

$^\dagger${\it Ruhr-Universit\"at Bochum, Institut f{\"u}r
  Theoretische Physik II,\\ D-44870 Bochum, Germany}\\

\bigskip

$^\star${\it Department of  Physics, University of Arizona, Tucson, Arizona 85721, USA}\\

\bigskip

$^\ast${\it Department of  Physics, Faculty of Engineering, Kyushu Institute of Technology, \\
1-1 Sensuicho, Tobata, Kitakyushu 804-8550, Japan}\\

\bigskip

$^\ddagger${\it Forschungszentrum J\"ulich, Institut f\"ur Kernphysik 
(Theorie),\\ D-52425 J\"ulich, Germany} \\
 and       \\
{\it Karl-Franzens-Universit\"at Graz, Institut f\"ur Theoretische Physik\\ 
A-8010  Graz, Austria}
\bigskip

$^\circ${\it  Jagiellonian University, Institute of Physics, Reymonta 4, \\
   30-059 Cracow, Poland} 

\bigskip

\end{center}

\vspace{.3in}

\thispagestyle{empty}

\begin{abstract}

\noindent
We perform the first 
complete analysis of $nd$ scattering at 
next--to--next--to--leading order  in 
chiral effective field theory including the corresponding three--nucleon force and  
extending our previous work, where 
only the two--nucleon  interaction has been taken into account. 
The three--nucleon force  appears first at this order in the chiral expansion 
and depends on two unknown parameters. 
These two  parameters are determined from the 
triton binding energy  and $nd$ doublet scattering length. We find an 
improved description of various scattering observables in relation to 
the next--to--leading order  results
especially at moderate energies 
($E_{\rm lab} = 65$ MeV). It is demonstrated that the long--standing $A_y$--problem
in  $nd$ elastic scattering is still not solved by the leading 3NF,
although some visible improvement is observed. 
We discuss possibilities of solving this puzzle. The predicted binding energy
for the $\alpha$-particle agrees with the empirical value.
\end{abstract}

\vfill

\pagebreak


\section{Introduction}
\def\theequation{\arabic{section}.\arabic{equation}}
\setcounter{equation}{0}

Effective field theory has become a standard tool for analyzing the
chiral structure  
of Quantum Chromodynamics (QCD) at low energy, where the perturbative
expansion in powers  
of the coupling constant cannot be used. The chiral symmetry of QCD is
spontaneously broken  
and the corresponding Goldstone bosons can be identified with
pions, if one considers the two flavor sector of the
up and down quarks as done here.
The pions are not exactly massless as it would be
the case for massless $u$ and $d$ quarks, but are much ligther than all other
hadrons and are therefore sometimes called Pseudo--Goldstone bosons.
It is a general property of Goldstone bosons that their interactions become weak for 
small momenta. Chiral Perturbation Theory (CHPT) is an effective field
theory  which allows 
to describe the interactions of pions and between pions and matter
fields (nucleons, $\rho$--mesons, $\Delta$--resonances, 
$\ldots$) in a systematic way. This is achieved via an expansion of
the scattering  
amplitude in powers of small external momenta and the pion mass. Pion
loops are naturally incorporated 
and all corresponding ultraviolet divergences can be absorbed at each
fixed order in the chiral expansion  
by counter terms of the most general chiral invariant
Lagrangian.  

This perturbative scheme works well in the pion and
pion--nucleon sector, where the interaction vanishes  
at vanishing external momenta in the chiral limit. The situation in
the purely nucleonic sector is  
somewhat different, since the interaction between nucleons is strong
and remains strong even in the chiral limit 
at vanishing 3--momenta of the external nucleons. The main difficulty
in the direct application 
of the standard methods of  CHPT to the nucleon--nucleon (NN) system
is due to the non-perturbative  
aspect of the problem. One way to deal with this difficulty has been
suggested by Weinberg, who  
proposed to apply CHPT to the kernel of the corresponding integral
equation for the scattering amplitude, 
which can be viewed as an effective NN potential \cite{weinberg90,weinberg91}.
  
Following this idea Weinberg was able to demonstrate  
the validity of the well--established intuitive hierarchy of the
few--nucleon forces: the two--nucleon 
interactions are more important than the three--nucleon ones, which
are more important than the four--nucleon 
interactions and so on. 

The first quantitative realization of the above idea has been
performed by  Ord\'o\~nez and co--workers, 
who derived the 2N potential and performed a numerical analysis of the
two--nucleon system \cite{ordonez96}.    
To calculate an expression for the effective Hamiltonian for two 
nucleons the authors of \cite{ordonez96} made use of   
Rayleigh--Schr\"odinger perturbation theory (the method is closely
related to the  
Tamm--Dancoff approach \cite{tamm45,dancoff50}),
which leads to a non--hermitian and energy--dependent potential. The
$\Delta$--degree of freedom has been included 
explicitely. The 26 free parameters, many of them being redundant due
to the property of  
antisymmetry of the wave functions, have been fixed from a global fit
to the low--energy observables.   
Ord\'o\~nez et al. obtained qualitative fits to deuteron properties as
well as quantitative fits to  
most of the scattering phase shifts up to $E_{\rm lab} =100$ MeV.

The property of the effective potential from Ref.~\cite{ordonez96} of
being explicitely energy--dependent  
makes it difficult to apply it to systems different from the
two--nucleon one. In fact, such an energy dependence  
is not a fundamental feature of the effective interaction and can be
eliminated by certain techniques,  
see e.g. \cite{friar99b}. In \cite{epelbaoum98b} we have demonstrated how to
derive the energy--independent 
and hermitian potential from the chiral Lagrangian using the method of
unitary transformation \cite{okubo54}. 
The advantage of this scheme is that it is easily extendable to
processes with more than two nucleons and/or  
external fields. In \cite{epelbaum00} we applied the above mentioned method
to calculate the NN scattering  
observables and deuteron properties up to NNLO in the chiral
expansion. As described in Ref. \cite{epelbaum02a}, the 9 
unknown low--energy constants (LECs)
related to contact interactions and the LECs $c_3$ and $c_4$ related 
to the subleading $\pi \pi$NN vertices
have been fixed by  a fit to the
Nijmegen phase shifts \cite{stoks94} in  
the $^1S_0$, $^3S_1-^3D_1$, $^1P_1$, $^3P_0$, $^3P_1$ $^3P_2-^3F_2$
channels below $E_{\rm lab}=100$ MeV. 
In contrast to Ref.~\cite{ordonez96} we did not perform a global fit to the data,
which due to the large dimension  
of the parameter space and computational resource limitations 
might not lead to the true global minimum in the $\chi^2$--space
and cannot easily be
performed.  Instead we introduced an alternative set 
of partial--wave projected LECs and considered each of the 
above indicated channels separately having at most 3 unknown parameters
in any given partial wave. 
The chiral potential at NNLO has been shown to lead to a reasonably
good description of the NN phase shifts  up to 
$E_{\rm lab} \sim 200$ MeV as well as of the deuteron
properties. Further we demonstrated that including  
the subleading two--pion exchange at NNLO allows to improve strongly 
the NLO results without introducing  
additional free parameters associated with short--range contact interactions, 
which is a good indication of consistency and convergence of the chiral 
expansion. For our choice of the LECs $c_{1,3,4}$ related to the subleading $\pi \pi NN$
interactions see \cite{epelbaum02a}. 

The few--nucleon interactions in chiral effective field theory have
been first discussed qualitatively  
by Weinberg \cite{weinberg91}. The corresponding expressions  have been
derived later by  
van Kolck, who demonstrated that the leading contribution, which
appears at NLO in the chiral expansion,  
cancels against the iteration of the energy--dependent part of the
corresponding NN effective potential  
\cite{kolck94}. Such a cancellation in case of the two--pion exchange
3NF has already been  observed earlier \cite{yang86}.
Thus the first non-vanishing contribution to the 3NF appears at
NNLO. Note that if the $\Delta$--resonance  
is included explicitely, i.e. if the $\Delta N$ mass splitting is
considered as a small quantity of the  
order of the pion mass, the non-vanishing contributions to the 3NF are
shifted to the NLO. 
Note, however, that such a scheme is not strictly rooted
in QCD because of the decoupling theorem \cite{gasser79}
(but can be justified in the large $N_c$--expansion, with $N_c$ the number of colors).   
For the energy--independent potential derived with the method of
unitary transformation one observes the  
vanishing of the NLO 3NF as well (as has also been pointed
out in a different context e.g.  in Ref. \cite{eden96}) 
and the first non-vanishing contributions appear at NNLO. 

In our work \cite{epelbaum01a} we performed a complete analysis of the
low--energy $nd$ scattering at NLO in the  
chiral expansion with the NN potential introduced in \cite{epelbaum00} and
also  calculated the  triton and $\alpha$--particle binding energies (BE's). 
Since no 3NF has to be included at this order and
all parameters in the NN potential are fixed from the 2N  
system, the results for $A>2$ systems are parameter--free
predictions. We demonstrated a reasonably good description of the  
$nd$ elastic scattering data at $E_{\rm lab} =3$ MeV and $E_{\rm lab}
=10$ MeV as well as of some break--up 
observables at  $E_{\rm lab} =13$ MeV 
while significant deviations from the data were found at $E_{\rm lab}
= 65$ MeV. The predicted value for the triton BE is in the
range comparable to the one based upon  
various modern phenomenological 
potentials, while for the $\alpha$--particle BE somewhat
larger deviations have been observed depending on the  
chosen cut--off value.  
Extending the analysis to NNLO requires, as already stated before, 
not only the appropriate modification of the NN interaction, but also 
the inclusion of the 3NF. In \cite{epelbaum02a} we presented an incomplete NNLO analysis of the 3N
system based upon the NN interaction at NNLO and without inclusion of
the 3NF.\footnote{From the point of view  
of an effective field theory it makes not much sense to include only
the 2N force and to omit the 3NF  
contributing at the same order. In \cite{epelbaum02a} we followed however
the common trend in the field of  
few--nucleon physics and calculated various 3N observables based on
our NNLO 2N potential for illustrative  
purposes.
Such comparisons are helpful to identify the observables and kinematics most
sensitive to the 3NF.
} 
In this work we present the complete NNLO analysis of the
low--energy $nd$ scattering  
including the chiral 3NF. 
We also predict the $\alpha$--particle binding energy.
This is the first time that the complete
chiral 3NF has been included in few--body  
calculations. Some pioneering steps in that direction based upon the hybrid
approach have been done in Ref.~\cite{huber01}.  

Our manuscript is organized as follows.
In  Section~II 
we discuss the structure of the chiral 3NF and
demonstrate that it depends  
on two parameters. The partial--wave decomposition of the
new terms in the 3NF  
is given in Appendix~A.
In Section~III we discuss how these unknown parameters can be fixed 
from {\em low-energy} 3N data. Then we show our results for various elastic and
break--up $nd$ scattering observables as  
well as for triton and $\alpha$--particle BE's in section~IV. 
Conclusions and an outlook are  given in Section~V.

\vfill\eject

\section{The Chiral Three--Nucleon--Force  at Next--to--Next--To--Leading Order}
\def\theequation{\arabic{section}.\arabic{equation}}
\label{sec:3NF}

\begin{figure}[htb]
\vspace{0.5cm}
\centerline{
\psfig{file=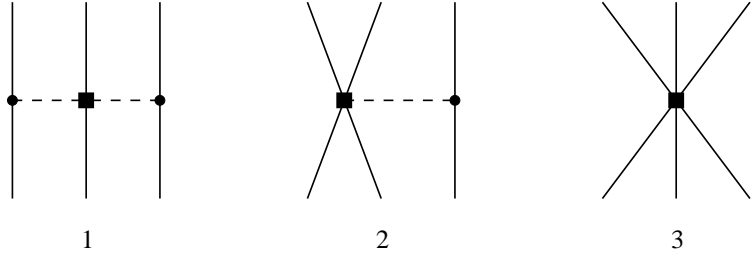,width=11cm}}
\vspace{0.3cm}
\centerline{\parbox{14cm}{
\caption[fig4]{\label{fig1} Three--nucleon force at NNLO: TPE, OPE and contact interaction.
Solid and dashed lines are nucleons and pions, respectively. Heavy dots denote leading vertices
with $\Delta_i=0$ and solid rectangles correspond to vertices of dimension $\Delta_i=1$. 
}}}
\vspace{0.5cm}
\end{figure}

The chiral 3NF at NNLO is given by the two--pion exchange
(TPE), one--pion exchange  
(OPE) with the pion emitted (or absorbed) by 2N contact interactions
and 3N contact interactions, 
see Fig.~\ref{fig1}. All diagrams include apart from the leading
vertices with $\Delta_i=0$ one insertion of 
interactions with $\Delta_i=1$, where the chiral dimension is defined
as
\beq
\Delta_i = d_i + \frac{1}{2} n_i - 2\,.
\eeq
Here $d_i$ and $n_i$ denote the number of derivatives (or pion mass insertions) and 
nucleon fields for a vertex of type $i$. This quantity has been  
first introduced by Weinberg and is especially useful in the few--nucleon sector. 
In the pion and pion--nucleon sectors one usually uses a different definition.

The contribution from the first graph in Fig.~\ref{fig1} is given (in
the 3N c.m.s.) by \cite{kolck94} (here and in what follows we use 
the usual notation for expressing the nuclear force: the quantity $V^{\rm 3NF}_{\rm TPE}$
is an operator with respect to spin and isospin quantum numbers and a matrix element with 
respect to momentum quantum numbers):
\beq
\label{3nftpe}
V^{\rm 3NF}_{\rm TPE}=\sum_{i \not= j \not= k} \frac{1}{2}\left(
  \frac{g_A}{2 f_\pi} \right)^2 \frac{( \vec \sigma_i \cdot \vec q_{i}
  ) 
(\vec \sigma_j \cdot \vec q_j  )}{(\vec q_i\, ^2 + M_\pi^2 ) ( \vec
q_j\, ^2 + M_\pi^2)}  F^{\alpha \beta}_{ijk} \tau_i^\alpha 
\tau_j^\beta \,,
\eeq
where  $\vec q_i \equiv \vec p_i \, ' - \vec p_i$; $\vec p_i$
($\vec p_i \, '$) are initial (final) momenta of the nucleon $i$ and 
\begin{displaymath}
F^{\alpha \beta}_{ijk} = \delta^{\alpha \beta} \left[ - \frac{4 c_1
    M_\pi^2}{f_\pi^2}  + \frac{2 c_3}{f_\pi^2}  
\vec q_i \cdot \vec q_j \right] + \sum_{\gamma} \frac{c_4}{f_\pi^2} \epsilon^{\alpha
\beta \gamma} \tau_k^\gamma  
\vec \sigma_k \cdot [ \vec q_i \times \vec q_j  ]\,.
\end{displaymath}
Here,
$g_A =1.276$ is the axial-vector coupling constant, $f_\pi = 92.4\,$MeV the weak
pion decay constant and the
 $c_{1,3,4}$ are the LECs from the chiral
Lagrangian of the  order $\Delta=1$  \cite{bernard95}, 
which also enter the corresponding expressions for
the subleading two--pion exchange in the  
2N potential. The form (\ref{3nftpe}) can be shown to match with the
low--momentum expansion 
of various existing phenomenological 3NFs provided they respect chiral
symmetry. This issue is extensively discussed 
in \cite{friar99a}.

We will now derive the expressions for the OPE and contact parts of the
3NF, see also \cite{kolck94}, and  
show that due to the Pauli principle only one independent OPE term and
one independent pure contact term  
appear in the 3NF.\footnote{Similar observation for the purely
  short--range part of the 3NF has been made  
by Bedaque et al. \cite{bedaque00}, while I.~Stewart pointed out that the
two OPE terms in the expressions for the 3NF   
published in \cite{kolck94} are not
independent from each other. Since these statements do not appear in
the literature in a complete form we decided  
to demonstrate this explicitly here.}

Let us start with the OPE contribution and discuss first the structure
of the corresponding $\pi$NNNN--vertex  
of dimension $\Delta=1$. After performing the non--relativistic
reduction for the nucleon field  (or, equivalently, after  
integrating out the lower components in the heavy--baryon
formalism) one encounters three different structures 
in the effective Lagrangian (in the rest--frame system of the nucleons):
\beq
\label{temp1}
\mathcal{L}^{(1)} = \alpha_1 ( N^\dagger N) (N^\dagger \vec
\sigma \fet \tau N ) \cdot  \vec \nabla \fet \pi +  
\alpha_2 ( N^\dagger \vec \sigma N) (N^\dagger \fet \tau N ) \cdot  \vec \nabla \fet \pi
+ \alpha_3 ( N^\dagger \vec \sigma \fet \tau N) \times (N^\dagger \vec
\sigma \fet \tau N ) \cdot  \vec \nabla \fet \pi\,. 
\eeq 
where $\fet \pi$ and $N$ denote the pion and nonrelativistic nucleon
fields, $\sigma_i$ and $\tau_i$  
are Pauli spin and isospin matrices. The symbol $\cdot$ ($\times$) denotes the simultaneous scalar
(vector) product in the ordinary and isospin--space. 
Note that the terms with derivatives acting on the nucleon fields are eliminated by partial integration.
The corresponding 3N force at NNLO is of the form
\beq
\label{temp2}
V^{\rm 3NF}_{\rm OPE} \propto \sum_{i \not= j \not= k} 
(\vec q_k \cdot \vec \sigma_k ) \ \frac{\vec q_k
  \fet \tau_k }{\vec q_k\, ^2 + M_\pi^2}  \cdot 
\biggl\{ 
\alpha_1 \, \vec \sigma_i \fet \tau_i + \alpha_2 \, \vec \sigma_i \fet \tau_j + \alpha_3 \,
(\vec \sigma_i \times \vec \sigma_j ) ( \fet \tau_i \times \fet \tau_j ) \biggr\} \,.
\eeq
Since we treat nucleons as identical particles,
the  few--nucleon states $| \Psi \rangle$ are antisymmetric. For these antisymmetric 
states the operators $V^{\rm 3NF}_{\rm OPE}$ can be 
further simplified. Because the force is symmetric 
with respect to an interchange of particles $i$ and $j$, the 
relation 
\begin{equation} 
V_{\rm OPE}^{\rm 3NF} \ | \Psi \rangle 
= {\mathcal A}_{ij} \  V_{\rm OPE}^{\rm 3NF} \ | \Psi \rangle =  V_{\rm OPE}^{\rm 3NF} \ 
{\mathcal A}_{ij} \ | \Psi \rangle
\end{equation}
holds and therefore one can work equally well with an antisymmetrized force.
Here $\mathcal{A}_{ij}$ is the antisymmetrization operator  in the space of two
nucleons $i$ and $j$, which reads: 
\beq
\mathcal{A}_{ij} = \frac{1-P_{ij}}{2}\,,
\eeq
where $P_{ij}$ is the corresponding permutation operator, $P_{ij} | ij \rangle = | ji \rangle$, given by 
\beq
P_{ij} = \frac{ 1 + \vec \sigma_i \cdot \vec \sigma_j}{2} \, \frac{1 + \fet \tau_i \cdot \fet \tau_j}{2}\,.
\eeq
In addition, one has to interchange the corresponding nucleon momenta. 
It is an easy exercise to apply the 
antisymmetrization operator $\mathcal{A}_{ij}$ to that pair $ij$ of the 3NF in eq.~(\ref{temp2})
which interacts via the contact terms and to see 
that all three different structures lead to the same expression. 

In the case of the purely contact 3NF without derivatives we proceed
in an analogous way. 
The most general structure of such 3NF which satisfies the usual
symmetry requirements (rotational and isospin  
invariance, parity invariance and invariance under time reversal
transformation) is given by 
\beqa
\label{temp3}
V^{\rm 3NF}_{\rm cont} = \sum_{i \not= j \not= k} &&\bigg\{ \beta_1 +
\beta_2 \, \vec \sigma_i \cdot \vec \sigma_j  
+ \beta_3 \, \fet \tau_i \cdot \fet \tau_j + \beta_4 \, (\vec \sigma_i
\cdot \vec \sigma_j ) ( \fet \tau_i \cdot 
\fet \tau_j ) + \beta_5 \, (\vec \sigma_i \cdot \vec \sigma_j ) ( \fet \tau_j \cdot \fet \tau_k ) \\
&& {} + \beta_6 \,( [ \vec \sigma_i \times \vec \sigma_j ] \cdot \vec
\sigma_k ) ( [ \fet \tau_i \times \fet \tau_j] 
\cdot \fet \tau_k ) \bigg\} \,. \nonumber
\eeqa
The antisymmetrization operator $\mathcal{A}_{ijk}$ in the space of
three nucleons can be expressed as:  
\beq
\mathcal{A}_{ijk} = \frac{(1 + P_{ij} P_{jk} + P_{ik} P_{jk} )}{3} \, \frac{(1-P_{jk} )}{2} \,.
\eeq
Acting with the operator $\mathcal{A}_{ijk}$ on the 3NF in
eq.~(\ref{temp3}) and performing a straightforward,  
but somewhat tedious simplification one  ends up with a single
structure just as in the  
previously considered case. We thus have shown that it is sufficient
to consider only one OPE and one  
pure contact term in the chiral 3NF at NNLO,
since all other terms have due to the Pauli principle precisely the
same effect on the  
S--matrix. In what follows, we will use the following form for
these 3NF contributions: 
\beqa
\label{3nfrest}
V^{\rm 3NF}_{\rm OPE} &=& - \sum_{i \not= j \not= k} \frac{g_A}{8
  f_\pi^2} \, D \, \frac{\vec \sigma_j \cdot \vec q_j }{\vec q_j\, ^2
  + M_\pi^2}  
\, \left( \fet \tau_i \cdot \fet \tau_j \right) 
(\vec \sigma_i \cdot \vec q_j ) \,, \\
V^{\rm 3NF}_{\rm cont} &=& \frac{1}{2} \sum_{j \not= k}  E \, ( \fet \tau_j \cdot \fet \tau_k ) \,,
\nonumber
\eeqa
where $D$ and $E$ are the corresponding LECs from the Lagrangian of
dimension $\Delta=1$: 
\beq
\mathcal{L}^{(1)} = - \frac{D}{4 f_\pi} \, ( N^\dagger N ) \,
(N^\dagger \vec \sigma \fet \tau N) \cdot \vec \nabla \fet \pi 
- \frac{1}{2} \, E \, ( N^\dagger N) \, (N^\dagger \fet \tau N ) \cdot (N^\dagger \fet \tau N )\,,
\eeq
Note that dimensional scaling
arguments allow 
one to express the LECs $D$ and $E$ as \cite{friar97}
\beq
D = \frac{c_D}{f_\pi^2 \Lambda_\chi}\, , \quad \quad E = \frac{c_E}{f_\pi^4 \Lambda_\chi}\,,
\eeq
where $c_D$ and $c_E$ should be numbers of order one and
$\Lambda_\chi$ is the chiral symmetry breaking 
scale of the order of the $\rho$ meson mass.
Here and in what follows we use $\Lambda_\chi = 700$~MeV.  
It has been demonstrated in \cite{epelbaum02b} that all 
corresponding numbers for 2N contact interactions at NLO and NNLO
are natural for the cut--off values considered.
It should also be understood that a more precise analysis
of the naturalness would  
require also taking into account symmetry factors in the Lagrangian as
well as additional factors  
resulting from insertions of spin and isospin matrices.\footnote{Such
  factors can be  
calculated from expressions of the 3NF. For example,
the antisymmetrized  
expression of the third term in eq.~(\ref{temp3}) is 3 times smaller
than the one of the fourth term, 
which has two additional insertions of the Pauli spin matrices.}

\section{Fixing the Parameters of the Three--Nucleon--Force}
\def\theequation{\arabic{section}.\arabic{equation}}

We now proceed to fix the unknown LECs $c_D$ and $c_E$ from 3N low-energy
observables. 
To that aim we solve the 3N Faddeev equations for the bound state
and for $nd$ scattering. They have the well known form \cite{glocklefb,nogga97}
\beq
\label{eq1}
\psi = G_0 \, t \, P \, \psi + (1 + G_0\, t) \,G_0 \,V_{\rm 3NF}^{(1)}
\,(1 + P)\, \psi\,, 
\eeq
in case of the bound state. Here $V_{\rm 3NF}^{(1)}$ is 
that part of the three-nucleon force
which singles out one particle (here particle 1)
and which is symmetrical under the exchange of the other
two particles. The complete 3NF is decomposed as
\beq
\label{decomp} 
V_{\rm 3NF} = V_{\rm 3NF}^{(1)}+ V_{\rm
  3NF}^{(2)}+V_{\rm 3NF}^{(3)}\,.
\eeq
Further, $\psi$ denotes the corresponding Faddeev component, $t$ is
the two--body $t$--operator, 
$G_0=1/(E-H_0)$ is the free propagator of three nucleons and $P$ is a
sum of a cyclical and anticyclical permutation 
of the three particles. In case of $nd$ scattering we follow our
by now standard path \cite{glockle96,huber97a} and firstly calculate a quantity $T$
related to the 
3N break-up process via the Faddeev--like equation:
\beq
\label{eq2}
T = t \,  P  \, \phi + (1 + t  \, G_0)  \, V_{\rm 3NF}^{(1)}  \, (1 +
P)  \, \phi + t  \, P  \, G_0  \, T + (1 + t  \, G_0)   
\, V_{\rm 3NF}^{(1)} \,  (1 + P)  \, G_0  \, T\,,
\eeq
where the initial state $\phi$ is composed of a deuteron and a
momentum eigenstate of the projectile nucleon.  
The elastic $nd$ scattering operator is then obtained as
\beq
U=P  \, G_0^{-1} + P  \, T + V_{\rm 3NF}^{(1)} \,  (1 + P) \,  (1 + G_0  \, T)\,,
\eeq
and the break-up operator via
\beq
U_0 = (1 + P)  \, T\,.
\eeq
These equations are accurately solved in momentum space using a
partial wave decomposition. For details see \cite{glocklefb,witala88,huber97b}. 
The corresponding partial wave decomposition of the chiral 3NF is given
in the appendix.
The equations (\ref{eq1}) and (\ref{eq2}) have to be regularized,
since the expressions 
for the 3NF (\ref{3nftpe}) and (\ref{3nfrest}) are only meaningful for
momenta below a certain scale. 
We regularize the $V^{\rm 3NF}$ in the way analogous to the one adopted in
the analysis of the two--nucleon system \cite{epelbaum00}:
\beq
V^{\rm 3NF} (\vec p, \vec q;\, \vec p\, ' , \vec q\, ') \rightarrow 
f_R (\vec p, \vec q) \, V^{\rm 3NF} (\vec p, \vec q;\, \vec p\, ' , \vec q\, ') \,
f_R (\vec p\, ', \vec q\, ')\,,
\eeq
where $\vec p$ and $\vec q$ ($\vec p\,'$ and $\vec q\,'$) are Jacobi momenta of the 
two--body subsystem and spectator nucleon before (after) the interaction. 
The regulator function 
$f_R (\vec p, \vec q \, )$ is chosen in the form 
\beq
f_R (\vec p, \vec q \, ) = \exp \left[ - 
\left(\frac{4 p^2 + 3 q^2}{4 \Lambda^2} \right)^2 \right]\,,
\eeq
so that it coincides with the exponential function $f_R^{\rm expon} (\vec p\, )$ of 
Ref.~\cite{epelbaum00} for $\vec q =0$. Clearly, this is not the
  only possible choice for that function. 
The final results for low--energy observables are insensitive to the choice of the 
regulator function provided that it does not violate the appropriate symmetries.
Note that the values of the LECs $c_D$ and $c_E$ \glqq{}run\grqq{} with the cut--off $\Lambda$
to compensate the changes in the observables, which are cut--off independent  
(up to the accuracy at the order in the chiral expansion). The dependence of the LECs on the cut--off $\Lambda$ 
is governed by renormalization group equations, as it is always the case in  
quantum field theory. We choose $\Lambda$ in the 3NF equal 
to $\Lambda$ in the NN interaction. The following study has been carried through with the
minimal and maximal momentum cut--offs, $\Lambda = 500$ and $600$ MeV, 
for which our NN force has been  defined in \cite{epelbaum02a}.  

The low--energy constants $c_D$ and $c_E$ 
enter the expressions for the chiral 3NF at NNLO.
The constant $c_E$ can only be obtained from 3N data, while $c_D$ can
be best determined in the 3N system or, for larger momentum
transfer, in pion production in NN collisions \cite{hanhart00}.
One important part of this work is to outline a feasible 
way to fix these parameters. We will now show 
that the LECs $c_D$ and $c_E$  can be determined using 
the $^3$H BE and the $nd$ doublet scattering length $^2a_{nd}$, which are
bona fide low--energy observables.
Since for the time being we have no $nn$ and $pp$ forces 
at our disposal (these have been calculated in chiral EFT
to NLO so far \cite{walzl01}) and both observables we are interested in are known 
to depend on  the difference between  $np$ and $nn$ forces, 
we decided to use $np$/$nn$ corrected data as input 
to our fitting procedure.  To this aim, we compare results 
using phenomenological forces with the proper $np$ and $nn$ 
forces and with a $np$ force only. 
Several combinations of NN
and 3N forces  have been adjusted to 
describe the triton BE (see \cite{nogga02b}).
We used AV18 \cite{wiringa95} augmented by 
the Urbana-IX 3NF \cite{pudliner97} and CD-Bonn 2000 \cite{machleidt01a}
augmented by the TM99' 3NF \cite{coon01}. 
These models come along with $nn$ forces, which 
are adjusted to the $nn$ scattering length. Replacing these $nn$ forces 
by the $np$ ones, we find an increased binding energy of 8.65 and 8.72~MeV,
respectively. From those we estimate a $np$ corrected 
 \glqq{}experimental\grqq{} pseudo BE of 
8.68~MeV.\footnote{In fact, it would also be sufficient 
to make an estimation of the isospin breaking effects based upon
purely 2N forces at the level of precision of NNLO.}

The \glqq{}experimental\grqq{} pseudo value for $^2a_{nd}$ has been  
determined using the NN force
CD-Bonn alone. The corresponding
shift of $^2a_{nd}$ is -0.19~fm. 
This together with the experimental value $^2a_{nd} = 0.64 \pm 0.04$ fm  
leads to the \glqq{}experimental\grqq{} pseudo value $^2a_{nd} =
0.45 \pm 0.04$. It should be  
understood that the uncertainty in the estimated pseudo value of the
scattering length is even larger 
due to the error in the shift resulting from replacement of the $nn$
force by the $np$ one. We however  
refrain from further discussion of that issue.

\begin{figure}[htb]
\vskip 1.0 true cm  
\begin{center}
\centerline{\hskip 0.2 true cm \psfig{file=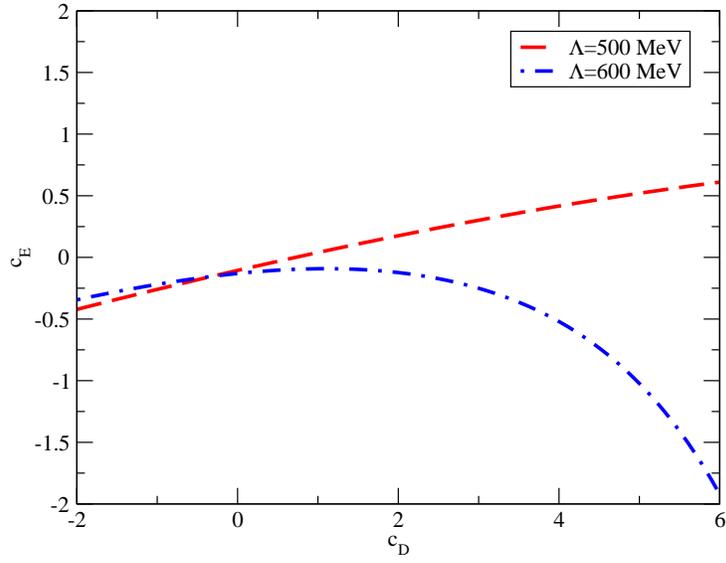,width=10cm}}
\centerline{\parbox{14cm}{
\caption{\label{fig2} Correlation between the LECs $c_E$ and $c_D$ after
  adjustment to the triton pseudo BE.}}} 
\end{center}
\end{figure}

For the chiral interactions at NNLO two unknown LECs enter 
into the 3N bound state Faddeev equation:
$c_D$ and $c_E$. Both affect the BE strongly. 
Imposing the condition that the Hamiltonian describes 
the pseudo BE, we find a
correlation between both LECs, which is  displayed in Fig.~\ref{fig2}. 
The unsymmetric interval shown in this figure is a consequence of the fact that 
the doublet scattering length favors positive values for $c_D$. 
The correlations have a very different behavior 
for both cut--off values. For $\Lambda=500$~MeV the functional 
form turns out to be nearly linear. 
This is not the case for $\Lambda=600$~MeV.
Later on we will demonstrate that this different behavior of the 
correlations for both cut--off values
does not show up in observables.  

\begin{figure}[hbt]
\vskip 1.0 true cm  
\begin{center}
\centerline{\hskip 0.2 true cm \psfig{file=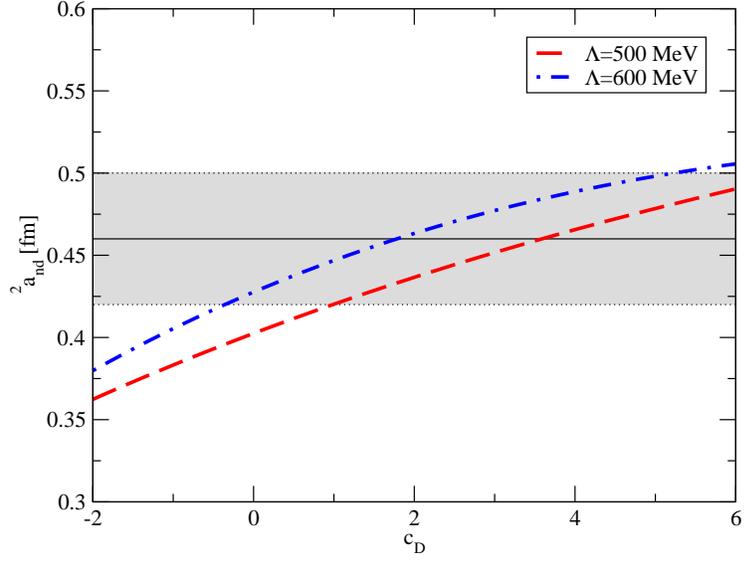,width=10cm}}
\caption{\label{fig3} $nd$ doublet scattering length $^2a_{nd}$ as function of the constant $c_D$.}
\end{center}
\end{figure}

One needs a second condition to fix both LECs uniquely.
The $nd$ doublet scattering length  $^2a_{nd}$ is 
known to be correlated  
with the $^3$H BE. 
This correlation is known as the Phillips line \cite{phillips68}.
We investigated it in the context of chiral nuclear
forces. It turns out that the scattering length 
depends on $c_D$ even if $c_E$ is chosen 
according to the correlation in Fig.~\ref{fig2} with the fixed value for the 
triton BE. This indicates that the correlation between the 
doublet scattering length  $^2a_{nd}$ and the $^3$H BE is not exact.
In fact, already for conventional NN and 3N forces,  
there was a slight scatter around an average line correlating the  $^3$H and $^2a_{nd}$
values for different nuclear forces\footnote{One should be careful by looking
at results which appear in the literature and sometimes indicate quite a strong 
deviation from the Phillips line. Especially earlier calculations 
have often been performed with not phase--equivalent potentials
and with restricted accuracy.} \cite{friar86b}. 
The Phillips line has recently been rediscovered within pion-less EFT \cite{bedaque99},\cite{bedaque02}.
At LO and NLO in the pion-less EFT the 3NF is given by a single contact term without 
derivatives and thus depends on just one free parameter. 
The Phillips line results from variation of this parameter and is in agreement with 
results based upon phenomenological interactions. Going to higher orders in the low--momentum 
expansion one encounters contributions to the 3NF with more derivatives and the exact 
correlation between $^2a_{nd}$ and $^3$H BE  observed at LO and NLO is broken, see \cite{bedaque02}
for more details. As discussed above, in the EFT with explicit pions the first nonvanishing 3NF 
at NNLO already depends on two free parameters and thus the Phillips line is already broken at this order in the 
chiral expansion. This allows to determine $c_D$ (and 
at the same time $c_E$) by a fit the the ``experimental'' pseudo datum for
the doublet scattering length.
In Fig.~\ref{fig3}  the grey horizontal
band indicates the scattering length range in agreement 
with the experimental error bar. Our theoretical predictions for $\Lambda = 500$ and $600$ MeV are
shown against $c_D$. 
We read off from Figs.~\ref{fig2}, \ref{fig3}   the following values:
\begin{eqnarray}
c_D &=& 3.6\, , {\quad} c_E = \phantom{-}0.37\, , \qquad \Lambda = 500~{\rm MeV}~,\nonumber \\
c_D &=& 1.8\, , {\quad} c_E =           -0.11\, , \qquad \Lambda = 600~{\rm MeV}~.
\end{eqnarray}
Notice that the sign of the determined LEC $c_D$ agrees
with the one found in \cite{hanhart00} from P--wave pion production in the 
proton--proton collisions. Note that for comparing our results with the ones 
of \cite{hanhart00} one should take into account different conventions with respect 
to $g_A$.
We are aware of the fact that we can only obtain a first estimate 
of $c_D$ and $c_E$. The most important uncertainties are the errors 
due to the $np$ force corrections and the experimental 
error bar of $^2a_{nd}$. In principle the errors due 
to these uncertainties with respect to observables
could be estimated 
performing calculations with several $c_D$ and $c_E$ combinations consistent 
with these error bars. In view of upcoming new data for $^2a_{nd}$ \cite{zimmerpriv,snowpriv}
and work on the isospin breaking 
in our formalism, we postpone such an analysis. In summary we emphasize
that the breakdown of the Phillips line correlation enables us to 
determine the LECs from the 3N BE and the $nd$ doublet scattering
length. The result is a parameter free 3N Hamiltonian. 
In the next sections we will investigate the 
results for the 4N bound state and 3N scattering
based on this Hamiltonian.

\section{Predictions for Three-- and Four--Nucleon Systems}
\def\theequation{\arabic{section}.\arabic{equation}}

We start with the prediction for the $\alpha$--particle 
BE. This is based on the solution of Yakubovsky equations 
\cite{yakubovsky67} as described in \cite{nogga00,nogga02b}.
The results are fully converged and accurate to 2~keV 
for the 3N and 20~keV for the 4N system. The convergence 
with respect to partial waves is much faster for the chiral 
interactions than for the convential ones.
This is a consequence of the momentum cut--offs, which suppress
the high momentum components exponentially. The calculations
of the binding energy 
for the chiral interactions are truncated at a two-body 
total angular momentum in the subsystem of $j_{\rm max}=6$ 
for the 3N system. For the 4N system we truncate the partial 
wave decomposition by the restriction that the sum of all 
three angular momentum quantum numbers is below $l_{\rm sum}^{\rm max}=10$. 
Calculations for conventional forces require   $l_{\rm sum}^{\rm max}=14$
(for details see \cite{nogga02b}).

Before we comment on our results for the BE's, we need 
to define a Coulomb and $np$ corrected $\alpha$--particle 
BE. 
Again, based on AV18+Urbana-IX and CD-Bonn+TM99',
we calculated BEs for the  $\alpha$--particle of 28.5~MeV 
and 28.4~MeV. Replacing the $pp$ and $nn$ forces by $np$ forces 
and omitting the Coulomb force, the BE's change to 29.9~MeV  
and 30.0~MeV. From these results we estimate an average 
change of the BE of $1.5\pm0.1$~MeV. The experimental 
$\alpha$--particle BE is 28.3~MeV. Thus we compare our 
results for the chiral interaction to an ``experimental'' 
pseudo BE of $29.8\pm0.1$~MeV.

In Table~\ref{tab1} this value is shown together with the ``experimental'' 
pseudo BE for $^3$H in comparison to the NLO and NNLO 
results for $^3$H and $^4$He. The BE is in general very sensitive 
to small changes of the interaction, as it comes out as the difference 
of the large kinetic and potential energies. As a consequence, we 
found a rather large dependence of the BE's on the cut--off 
at NLO \cite{epelbaum01a} ($\sim$19~\% for the $\alpha$--particle). 
At NNLO the $^3$H BE agrees with the ``experimental'' value
by construction. Because of the strong correlation of 3N 
and 4N BEs, known as Tjon-line \cite{tjon75}, one can expect 
a rather small cut--off dependence of the $\alpha$--particle BE, too. 
This is indeed the case. However, we would like 
to mention that 3NFs break this correlation \cite{nogga02b}
and we observe  a $c_D$ dependence of the $\alpha$--particle 
BE (1.5~MeV change in the range $c_D=-1.5 .. +1.5$ for $\Lambda=500$~MeV). 
We are also very encouraged by the fact that the $\alpha$--particle 
BE for both cut--offs comes out close to the ``experimental'' value. 
Note also that no 4NF contributes at NNLO. 
Therefore all predictions for $A>3$ at NNLO are parameter free.
 
\begin{table}[t]
\vskip 0.7 true cm
\begin{center}
\begin{tabular}{||l||r||r|r|r|r||r|r|r|r||}
\hline \hline
               &             & \multicolumn{4}{c||}{$^3$H} 
                             & \multicolumn{4}{c||}{$^4$He}  \\
\cline{3-10}
               &  $\Lambda$  &  $E$  &  $T$  &  $V_{NN}$ & $V_{3N}$
                             &  $E$  &  $T$  &  $V_{NN}$ & $V_{3N}$ \\
\hline \hline
NLO            &  500        & $-$8.54  & 30.76 & $-$39.30    & ---    
                             & $-$29.57 & 61.42 & $-$91.00    & --- \\
               &  600        & $-$7.53  & 39.24 & $-$46.77    & ---    
                             & $-$23.87 & 77.61 &$-$101.47    & --- \\
\hline
NNLO           &  500        & $-$8.68  & 31.07 & $-$39.43    & $-$0.318 
                             & $-$29.51 & 61.83 & $-$89.59    & $-$1.753 \\
               &  600        & $-$8.68  & 34.44 & $-$42.41    & $-$0.712
                             & $-$29.98 & 71.49 & $-$97.44    & $-$4.025 \\
\hline
``Expt''       &  ---        & $-$8.68  & ---   & ---       & ---   
                             & $-29.8\pm0.1$ & ---      &  ---         &  ---       \\
\hline \hline
\end{tabular}
\caption{$^3$H and $^4$He BE at NLO and NNLO of the chiral
  expansion 
(for the cut--off range considered throughout) compared to ``experimental''
pseudo BE (see text). Apart from the BE's $E$ (in MeV), we  also give  
the kinetic energies $T$ (in MeV) as well as expectation values of 2N
and 3N forces  
$V_{NN}$ and  $V_{3N}$, respectively,  (in MeV).
\label{tab1}}
\end{center}
\vskip 0.5 true cm
\end{table}

Additionally, we list the expectation values 
of the different parts of the Hamiltonian in Table~\ref{tab1}. 
It is important to realize that those quantities are not observable.
We see that the relative contributions of the NN and 3NF parts
are comparable in the 3N and 4N system. We also observe 
that the ratio of NN and 3NFs strongly depends on the  
cut--off chosen. This is elaborated in more detail in Table~\ref{tab1a}, where 
the 3NF expectation value is split in the contribution from 
the $2\pi$ exchange (c-terms), $1\pi$ exchange (D-term) and contact term
(E-term). The contributions of D- and E-term cancel to a large 
extent in both nuclei and for both cut--offs. The change 
in sign of the E-term changing from $\Lambda=500$~MeV to $\Lambda=600$~MeV 
has to be expected, since the $c_E$ changes its sign, too. 
More surprising is the change in sign for the D-term. This has to
be caused by a qualitatively different action of the D-term 
operator on the wave functions for $\Lambda=500$~MeV and $\Lambda=600$~MeV.

\begin{table}[t]
\vskip 0.7 true cm
\begin{center}
\begin{tabular}{||l||r||r|r|r|r||r|r|r|r||}
\hline \hline
               &             & \multicolumn{4}{c||}{$^3$H} 
                             & \multicolumn{4}{c||}{$^4$He}  \\
\cline{3-10}
               &  $\Lambda$  &  c-terms  &  D-term &  E-term & all 
                             &  c-terms  &  D-term &  E-term & all \\
\hline \hline
NNLO           &  500        & $-$0.39  &  0.81 & $-$0.74     & $-$0.32  
                             & $-$2.00  &  3.93 & $-$3.69     & $-$1.75  \\
               &  600        & $-$0.73  & $-$0.12 &  0.13     & $-$0.71 
                             & $-$3.81  & $-$0.84 &  0.63     & $-$4.03  \\
\hline \hline
\end{tabular}
\caption{Contribution of the different terms of the 3NF to 
         the complete 3NF expectation value for $^3$H and $^4$He. 
         All energies are given in MeV.
\label{tab1a}}
\end{center}
\vskip 0.5 true cm
\end{table}

We now switch to scattering observables.
Most of the 3N scattering data have been obtained for the $pd$ 
system. In the case of scattering the isospin breaking effects in the nuclear force
are believed to be of minor importance. We have checked this assumption
explicitly for elastic scattering observables using the CD Bonn potential 
with $np$ and $nn$ and with only 
$np$ forces to evaluate corresponding effects. Only in two cases at 3 MeV, namely 
for $T_{20}$ (at forward angles data are shifted upwards) and for $T_{21}$
(data are shifted downwards at angles below 120$^\circ$), significant effects were found.
The $np$--force corrections are small for all considered elastic scattering observables 
at 10~MeV and nearly invisible at 65~MeV.
Therefore we refrain from correcting data for this effect. 
In contrast, there are visible Coulomb corrections necessary at 
these energies. We are not able to take the Coulomb force into account
in the 3N continuum. For the Coulomb corrections 
we rely on the work of the Pisa collaboration, who can 
calculate low energy scattering observables based on the full AV18
interaction including the Coulomb force 
\cite{kievskypriv,kievsky99}. The difference of these 
full calculations and calculations without Coulomb force
serves as our estimate of the Coulomb corrections. 
In the following, all $pd$ elastic scattering data 
at 3~MeV and 10~MeV have been corrected by this amount. 
For 65~MeV we did not correct the data assuming that Coulomb corrections are 
small except in forward direction. For the break--up we refrained 
from any corrections because of the lack of reliable theoretical calculations 
taking the Coulomb force into account.
  
The $nd$ scattering  observables have been studied very intensively
using the modern 
phenomenological interactions \cite{glockle96,kievsky99}.
In general, the description of the data by these models 
is very good at low energies with a few prominent exceptions. The most 
famous one entered the literature  as $A_y$ puzzle \cite{koike87,witala94}, which is 
related to the fact that this observable is underpredicted 
in the maximum by realistic high--precision models of the nucleon interactions.
In this paper we do not compare the new results to 
traditional ones. That has been done in \cite{epelbaum02a} 
for the NLO and NNLO interactions without 3NF part. 
In this case, the NNLO interaction 
compares quite well to the results \cite{witala01a}   based 
on the highly accurate phenomenological forces.
In the following we would like to concentrate on 
the {\em complete}  analysis at NNLO.

\begin{figure}[hbt]
\vskip 0.0 true cm  
\begin{center}
\centerline{\hskip 0.2 true cm \psfig{file=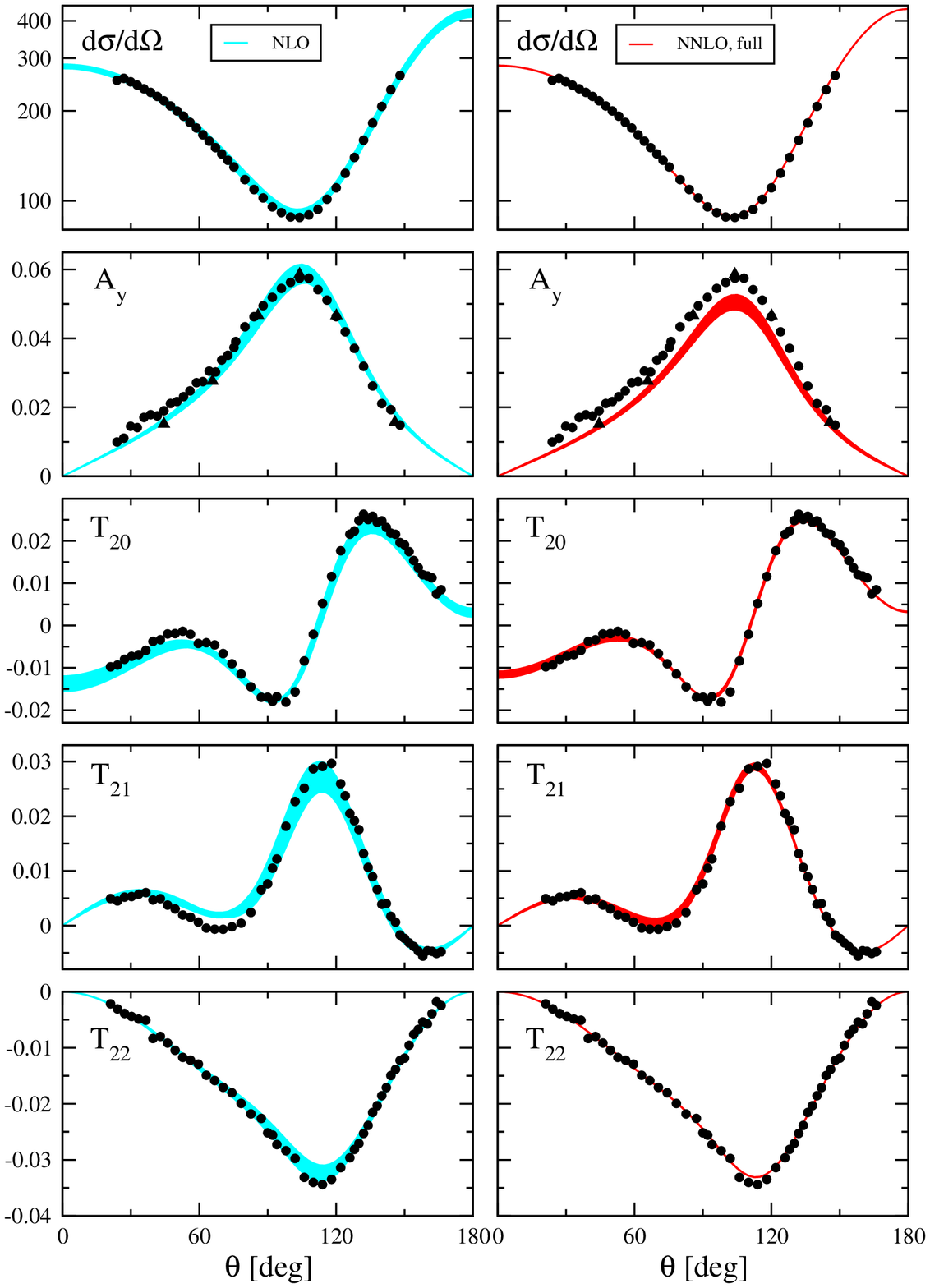,width=14cm}}
\centerline{\parbox{14cm}{
\caption{\label{fig4} $nd$ elastic scattering observables at 3 MeV at NLO (left column) and
NNLO (right column). The filled circles are $nd$ pseudo data based on \cite{shimizu95,sagara94} while the 
filled triangles are true $nd$ data \cite{mcaninch93}. The bands correspond 
to the cut--off variation between 500 and 600 MeV. The unit of the cross section is mb/sr.}}}
\end{center}
\end{figure}

\begin{figure}[hbt]
\vskip 0.0 true cm  
\begin{center}
\centerline{\hskip 0.2 true cm \psfig{file=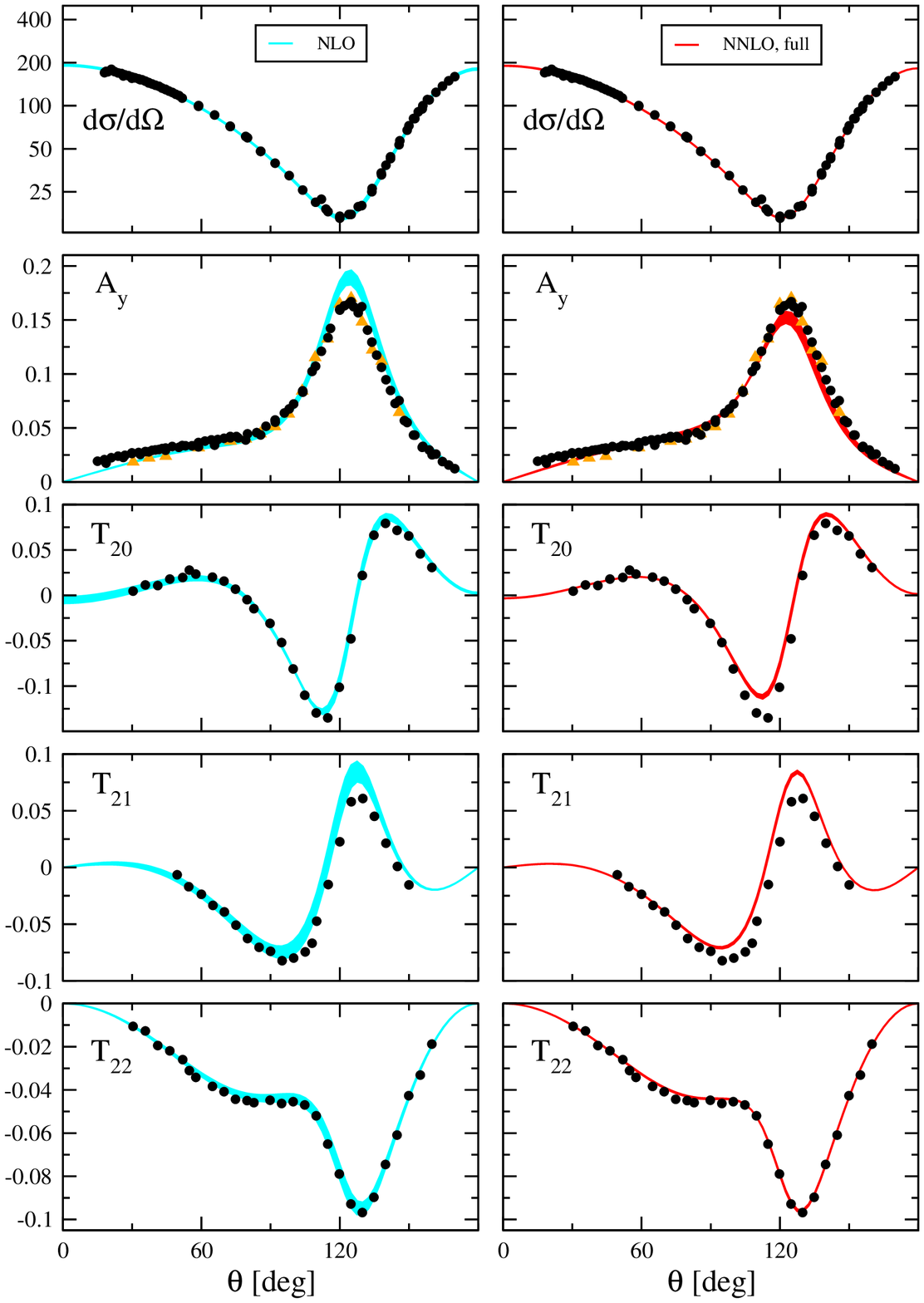,width=14cm}}
\centerline{\parbox{14cm}{
\caption{\label{fig5} $nd$ elastic scattering observables at 10 MeV at NLO (left column) and
NNLO (right column). The filled circles are $nd$ pseudo data based on  \cite{rauprich88,sperisen84,sagara94} 
while the filled triangles are true $nd$ data \cite{howell87}. The bands correspond 
to the cut--off variation between 500 and 600 MeV. The unit of the cross section is mb/sr. }}}
\end{center}
\end{figure}

\begin{figure}[hbt]
\vskip 0.0 true cm  
\begin{center}
\centerline{\hskip 0.2 true cm \psfig{file=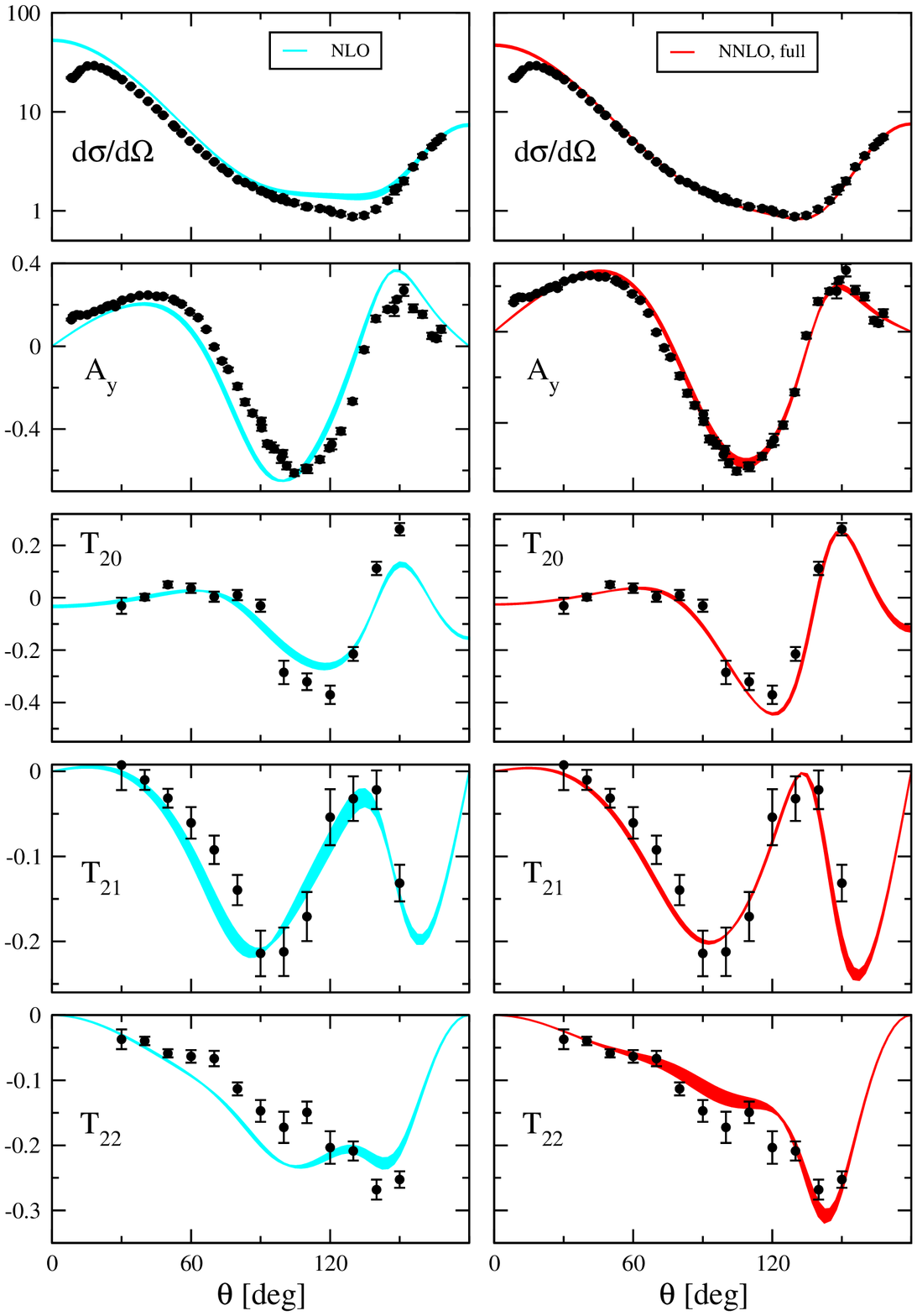,width=14cm}}
\centerline{\parbox{14cm}{
\caption{\label{fig6} $nd$ elastic scattering observables at 65 MeV at NLO (left column) and
NNLO (right column). The filled circles are $pd$  data \cite{shimizu95,witala93}. 
The bands correspond to the cut--off variation between 500 and 600 MeV.  The unit of the cross section is mb/sr.}}}
\end{center}
\end{figure}

In Figs.~\ref{fig4}, \ref{fig5} and \ref{fig6} we show a 
comparison for few selected elastic scattering observables
at 3, 10 and 65~MeV, respectively. 
The left column shows our results for NLO \cite{epelbaum01b}
in comparison to the data and the right column the new NNLO 
results compared to the same data. 
The bands are given by the cut--off variation in the range from $\Lambda=500$ 
and $\Lambda=600$~MeV. They may serve as an estimation of the effects
due to neglected higher order terms in the chiral expansion.
 
\begin{figure}[hbt]
\vskip 0.0 true cm  
\begin{center}
\centerline{\hskip 0.2 true cm \psfig{file=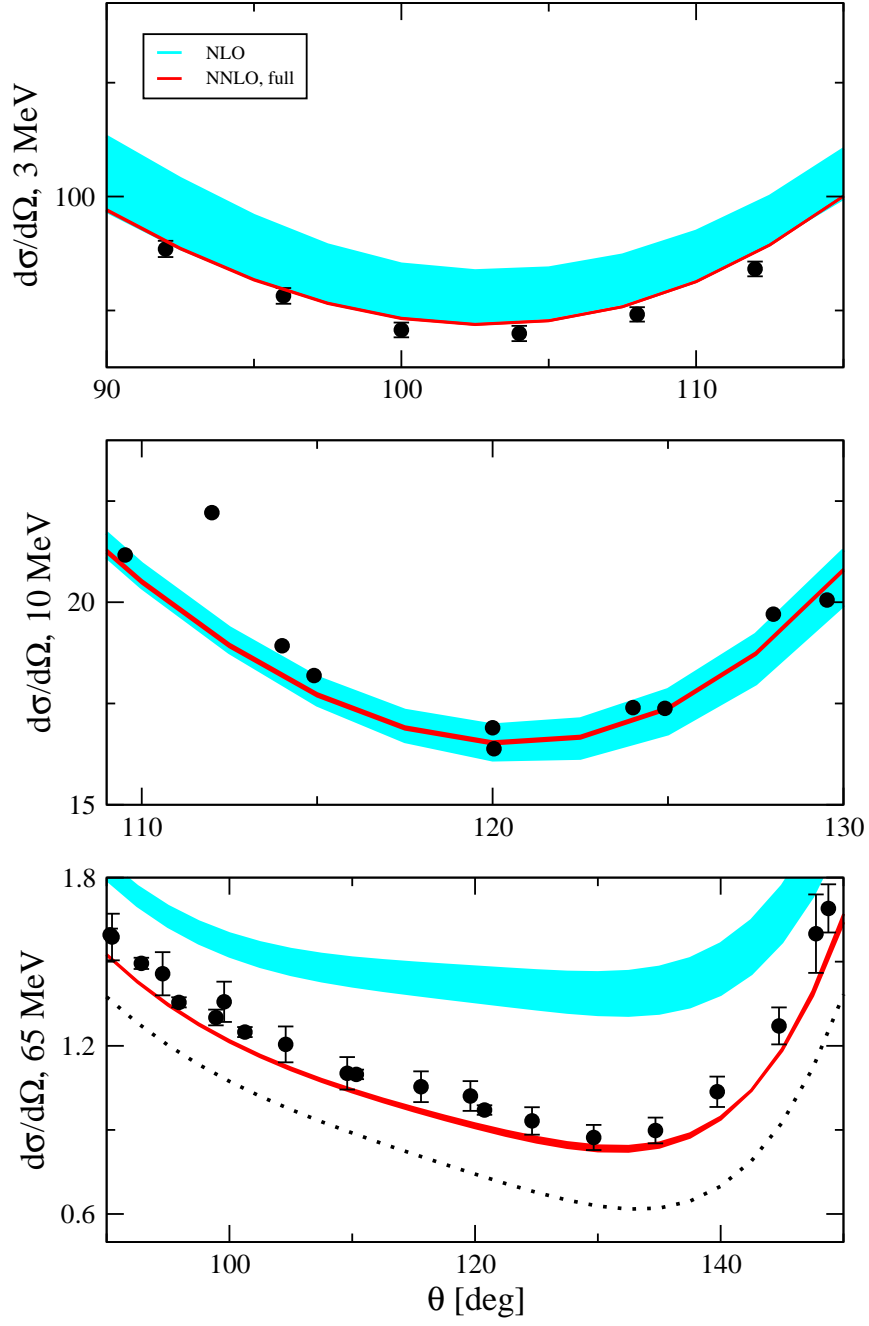,width=12cm}}
\centerline{\parbox{14cm}{
\caption{\label{fig6a} Minima of the cross section (in mb/sr) of elastic $nd$ scattering at 3 MeV 
(upper panel), 10 MeV (panel in the middle) and 65 MeV (lower panel) at NLO and
NNLO. The filled circles are $nd$ pseudo data
at 3 and 10 MeV and true $pd$ data at 65 MeV. 
The bands correspond to the cut--off variation between 500 and 600 MeV.
The dotted line at 65 MeV shows  the NNLO result with $c_D = -3.0$ and $\Lambda = 500$ MeV.
}}}
\end{center}
\end{figure}

The differential cross section is presented in the first 
row of Figs.~\ref{fig4}, \ref{fig5} and \ref{fig6}. Additionally,
we give a more detailed look at the cross section minima in Fig.~\ref{fig6a}.
At 3~MeV and 10~MeV we see that NLO and  
NNLO predictions overlap. The cut--off dependence is already
small at NLO and nearly vanishes at NNLO. This strong reduction
of the cut--off dependence of this observable at NNLO is expected 
and can easily be understood. Indeed, at least at low energy the 
differential cross section is dominated by the 2N S--waves. 
The situation is more interesting at 
65~MeV. In the minimum of the cross section 
one observes large differences between the NLO
and NNLO results (also to the incomplete NNLO calculation, see \cite{epelbaum02a}). 
The cut--off dependence of the NLO 
results is more visible than at lower energies,
and is again strongly reduced at NNLO. 
The NNLO results are in agreement with the data
except  for forward directions, which 
are sensitive to the Coulomb force. 
Note that the improvement at NNLO is not only due to the fact that 
the NNLO 2N potential leads to a much more accurate description of the 
data especially at moderate energies \cite{epelbaum02a}, but also due to the chiral 3NF. 
This is demonstrated by the 
dotted line in the lower panel of figure \ref{fig6a}, which  
corresponds to  $c_D=-3.0$ at $\Lambda=500$~MeV  
(and a $c_E$ chosen appropriately to reproduce the 3N binding energy).\footnote{This 
value of  $c_D$ is excluded by the observed value of the doublet scattering length.}
For this value of $c_D$ the prediction in the minimum is in 
disagreement with the data. It is gratifying to see that fixing the LEC $c_D$ 
from the scattering data at zero energy we are able to describe the cross section 
minimum at 65 MeV.
We consider 
this to be an important indication of consistency in the determination of
the LECs $c_D$ and $c_E$.

As already pointed out before,
the most problematic observable of $nd$ elastic scattering is 
$A_y$, which is shown in the second row of  
Figs.~\ref{fig4}, \ref{fig5} and \ref{fig6}.  
First of all we would like to stress that vector and 
tensor analysing powers are defined as differences 
of polarized cross sections and are rather small at low energies,
so that larger theoretical errors for these observables have to be expected.
At energies 3 and 10 MeV we see visible 
deviations of our predictions for $A_y$ from the data for both NLO and  NNLO. 
It is well known \cite{kievsky98} that this observable is extremely sensitive 
to the $^3$P--wave phase shifts in the NN system. Although at NLO chiral predictions at 3 MeV 
seem to be in agreement with the data, this cannot be considered as a final result 
in chiral EFT. Indeed, the $^3$P--wave phase shifts in the NN system
are only described at low energies with an accuracy of about 5$\%$ \cite{epelbaum02a}, which 
indicates that large corrections to $nd$ $A_y$ at higher orders in the chiral expansion are possible.
At NNLO the $^3$P--wave phase shifts come out with a significantly smaller 
error of about 2$\%$ (at $T_{\rm lab} = 10$ MeV) and we found in \cite{epelbaum02a} that  
$A_y$ is underpredicted if one only uses the 2N forces, just as in the case of high--precision 
potential models. As one sees from Figs.~\ref{fig4} and \ref{fig5}, we do not solve the 
$A_y$ puzzle performing the complete NNLO analysis and including the 3NF. 
It is important to stress 
that in principle, one could try to solve this puzzle by the NNLO chiral 3NF. Indeed, 
instead of fixing the unknown LECs $c_D$ and $c_E$ to the triton binding energy and the doublet
scattering length, one could think about requiring a good description of $A_y$ at, say, 3 MeV
as being one of the two conditions needed. We found, however, that 
$A_y$ is not very sensitive to the choice of the LECs $c_D$ and $c_E$,
if the two are adjusted to reproduce the triton binding energy.  
We were not able to find values of these coefficients in the natural range, which would 
simultaneously describe $A_y$ at 3 MeV and the triton BE.\footnote{H\"uber et al.~pointed 
out in \cite{huber01}, that  $A_y$ is sensitive to the choice of $c_D$. This is not in contradiction with 
the above discussion. We also observed a sensitivity to independent variations in $c_D$ and $c_E$,
which is, however, strongly reduced as soon as one requires to reproduce the triton BE.}
This indicates that higher order effects are probably still important for 
this observable. On the other hand it can be also a hint that the $A_y$ puzzle
is related to an insufficient knowledge of the low energy $^3P_j$ NN phases \cite{tornow98}.
It will be interesting, as a next step,  to include $1/m_N$ corrections 
to the interactions and to study their effect 
especially on $A_y$. At 65~MeV the decription of $A_y$  
is much better, which is in agreement with results based upon the 
phenomenological nuclear forces. 

The lower three rows of Figs.~\ref{fig4}, \ref{fig5} and \ref{fig6} show 
the tensor analyzing powers $T_{20}$, $T_{21}$ and $T_{22}$. 
At 3~MeV and 10~MeV, we find in general that the 
NNLO predictions stay within the NLO band.
The cut--off dependence clearly shrinks, which is a good indication of 
convergence of the chiral expansion.
Alltogether the agreement with the data is good 
except for $T_{20}$ and $T_{21}$ at 10~MeV.\footnote{Significant deviation from the data 
in the minimum of $T_{21}$ can be observed at 3 MeV as well, 
if $np$--force corrections are taken into account.}
Notice that similar results have been reported in \cite{Kievsky01} based 
upon the combination of the AV18 2N and the Urbana~IX 3N forces, where these observables
have been calculated in the $pd$ system and the Coulomb force has been taken 
into account. 
At 65~MeV, the situation is comparable to the one for 
the cross section. While the NLO predictions at this energy deviate significantly 
from the data, the NNLO results are in a much better agreement. 
Unfortunately, the quality of the data does not allow to draw more 
precise conclusions and especially here new high--precision data are needed.

\begin{figure}[hbt]
\vskip 0.0 true cm  
\begin{center}
\centerline{\hskip 0.2 true cm \psfig{file=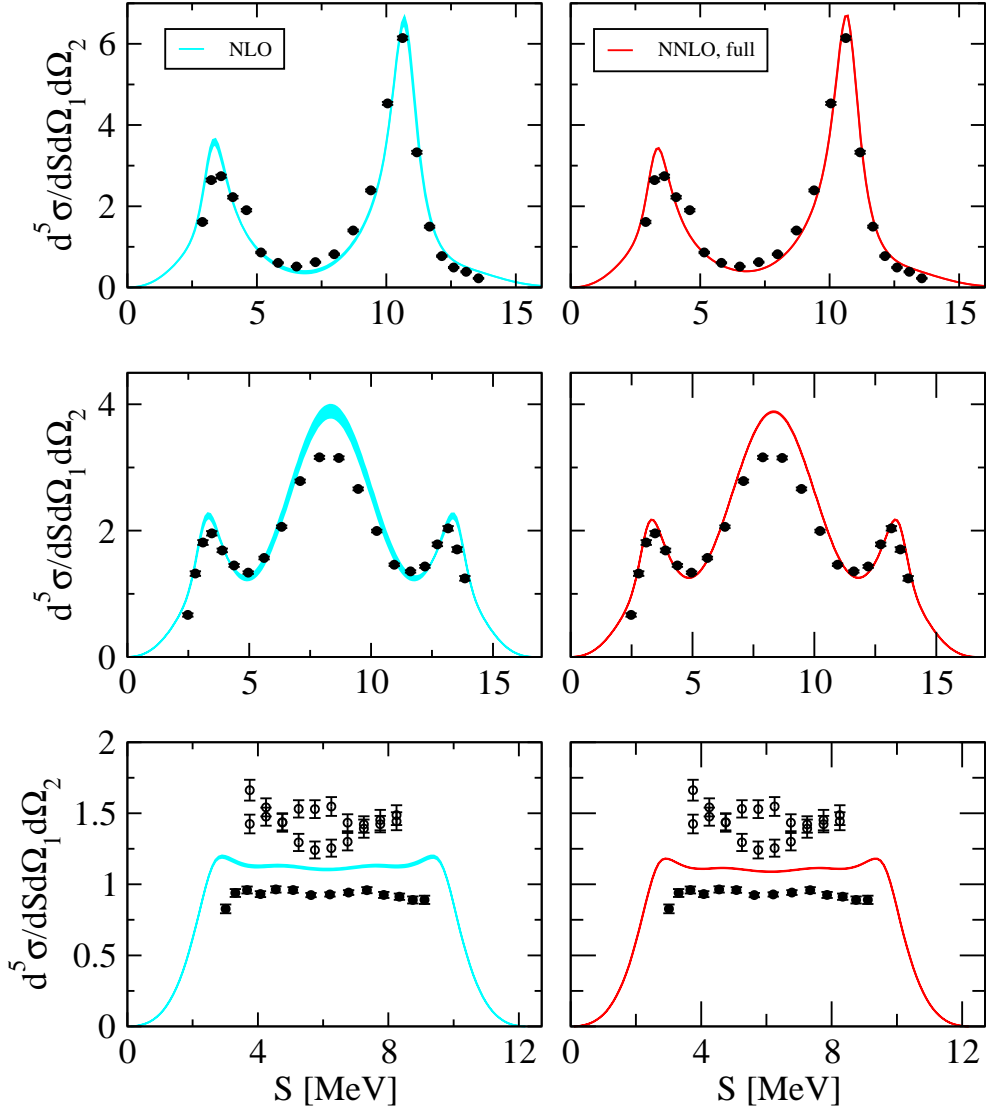,width=14cm}}
\centerline{\parbox{14cm}{
\caption{\label{fig7} $nd$ break up cross section in [mb MeV$^{-1}$ sr$^{-2}$] along the kinematical locus $S$ (in MeV) 
at 13 MeV in comparison to predictions at NLO (light shaded band) and NNLO (dark shaded band) in chiral effective
field theory. In the upper row a final state interaction configuration is shown, in the middle one 
a quasi--free scattering configuration (both in comparison to $pd$ data) 
and in the lower one a space star configuration (upper data $nd$, lower data $pd$). 
The precise kinematical 
description can be found in Ref.~\cite{glockle96}. 
$pd$ data are from \cite{rauprich91}, $nd$ data from \cite{setze96,strate89}. }}}
\end{center}
\end{figure}

Let us now switch to break--up observables.
We performed calculations at two energies, 13 and 65~MeV, where 
a lot of $pd$ data exist. 
As already pointed out above, there are no reliable Coulomb corrections available for 
the break up. 
Therefore we show the non--corrected $pd$ data 
in comparison to our $nd$ calculations.  
Note that for the space star configuration at 13~MeV presented in Fig.~\ref{fig7}, 
it is shown that the $nd$ and $pd$ cross section data strongly deviate 
indicating that Coulomb effects 
can become important at least in some configurations.
Presumably, the Coulomb corrections are smaller at 65~MeV.

At 13 MeV we demonstrate in Fig.~\ref{fig7} 
chiral predictions for the cross section in the often investigated final--state 
interaction peak, quasi--free scattering and space--star configurations, 
which have also been considered in the NLO analysis \cite{epelbaum01a}.
For a general discussion on various break up observables and configurations the 
reader is referredd to Ref.~\cite {glockle96}.
As demonstrated in Fig.~\ref{fig7}, the NLO and NNLO 
results essentially agree at 13 MeV. They describe the
configuration dominated by FSI peaks quite well (for a more
elaborated procedure the angular openings of the detectors have
to be taken into account, see \cite{glockle96}). 
The present theory for the break--up
configuration including a QFS geometry fails in the central maximum.
The reason might be Coulomb force effects. The third
configuration, the so called space--star, is one of the long standing 
puzzles of 3N scattering \cite{kuros02a,kuros02b,howell98a,zhou01,tachikawa01}.
Similar to the phenomenological interactions, we even fail to describe 
the $nd$ data. $pd$ and $nd$ data are quite different and it remains
open whether  Coulomb corrected data would fall on
the theory. We observe \cite{glockle96} the tendency that with conventional NN forces 
theory is already rather close to the $pd$ data at 19~MeV and even closer at 65~MeV. This suggests
that presumably the discrepancy to $pd$ data is due to Coulomb force effects.

\begin{figure}[hbt]
\vskip 0.0 true cm  
\begin{center}
\centerline{\hskip 0.2 true cm \psfig{file=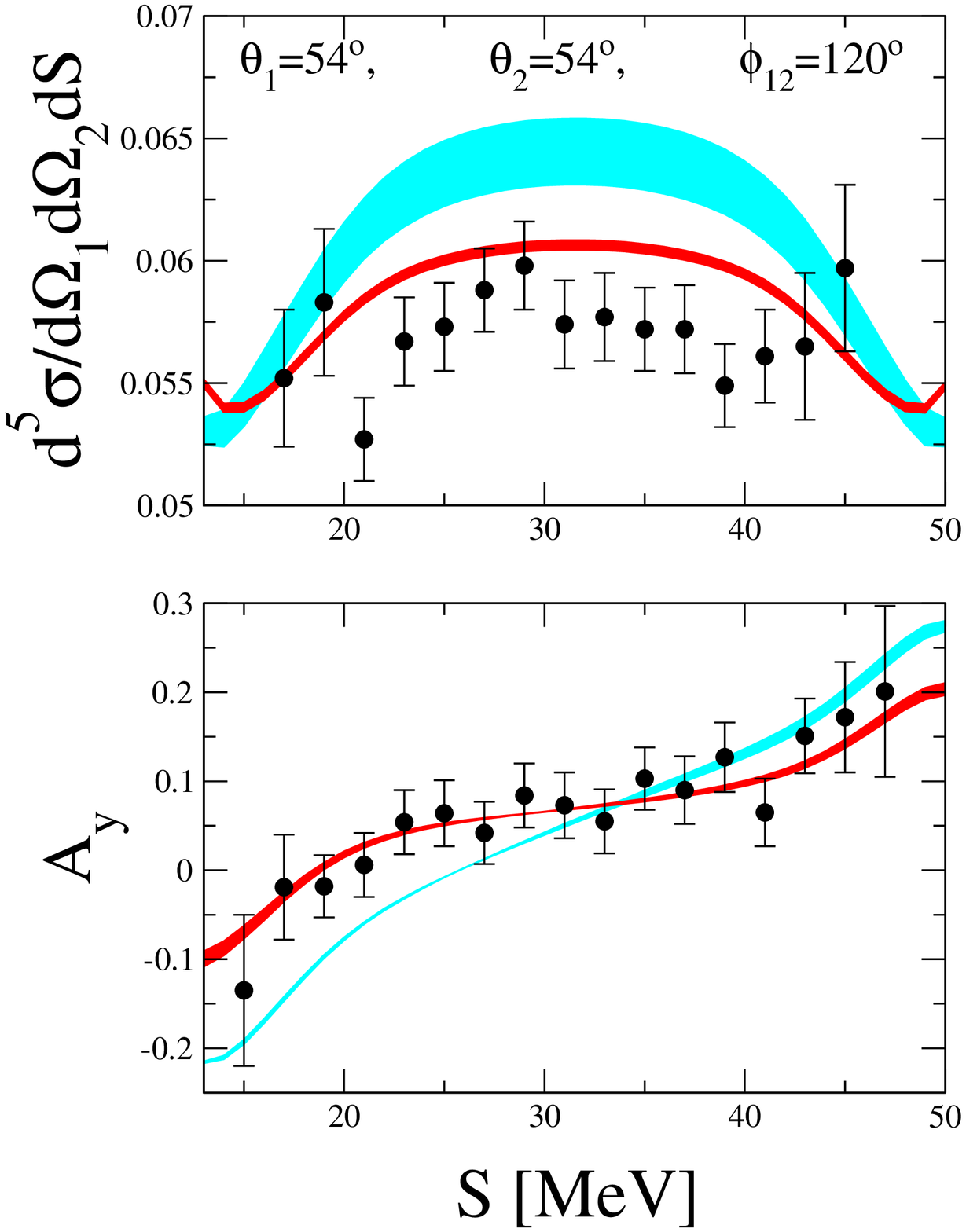,width=14cm}}
\centerline{\parbox{14cm}{
\caption{\label{fig8} $nd$ break up cross section in [mb MeV$^{-1}$ sr$^{-2}$] and nucleon analyzing power 
along the kinematical locus $S$ (in MeV) 
at 65 MeV in comparison to predictions at NLO (light shaded band) and NNLO (dark shaded band) in chiral effective
field theory. Symmetric space star configuration is shown. $pd$ data are from \cite{zejma97}}}}
\end{center}
\end{figure}

\begin{figure}[hbt]
\vskip 0.0 true cm  
\begin{center}
\centerline{\hskip 0.2 true cm \psfig{file=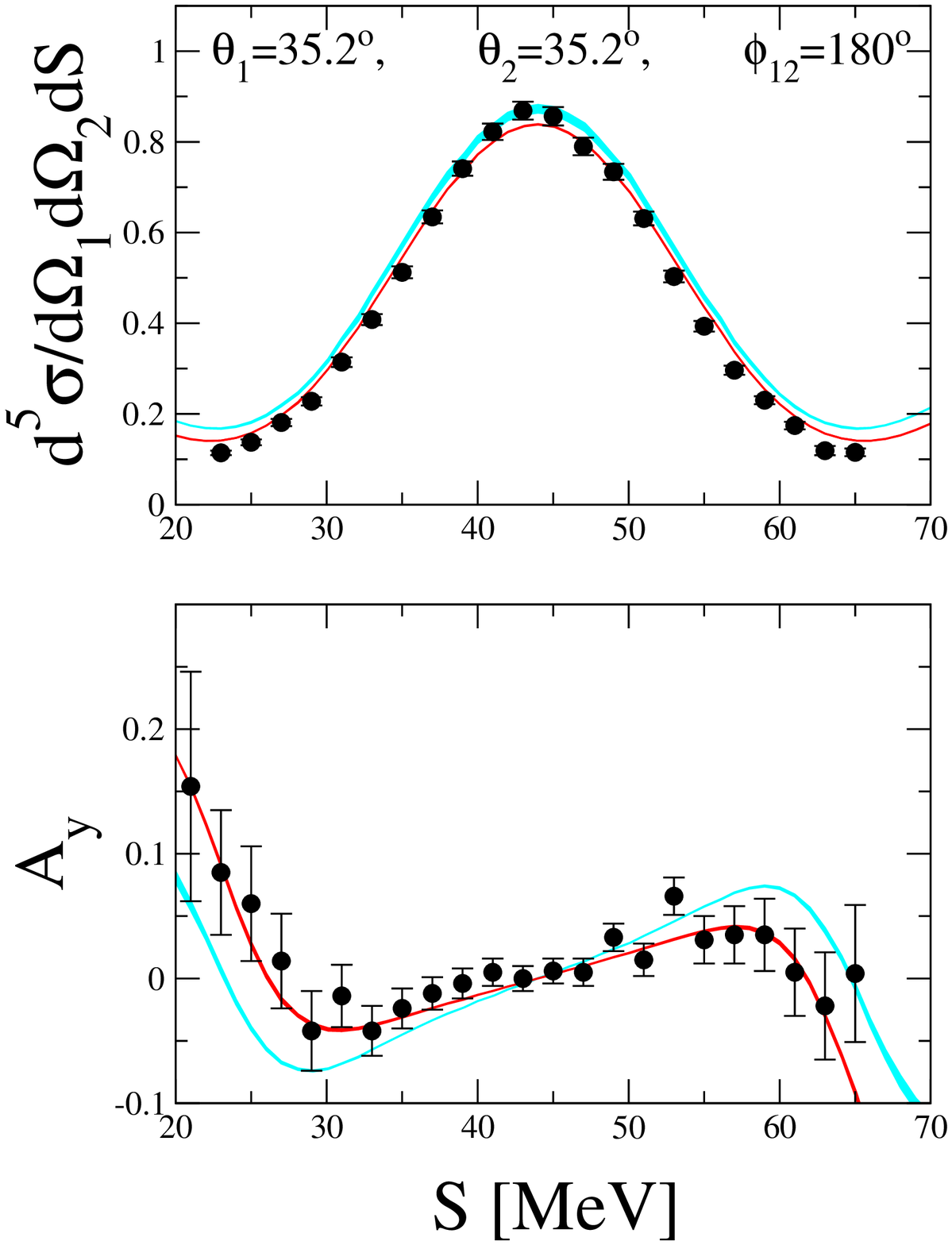,width=14cm}}
\centerline{\parbox{14cm}{
\caption{\label{fig9} $nd$ break up cross section in [mb MeV$^{-1}$ sr$^{-2}$] and nucleon analyzing power 
along the kinematical locus $S$ (in MeV) 
at 65 MeV in comparison to predictions at NLO (light shaded band) and NNLO (dark shaded band) in chiral effective
field theory. Symmetric forward star configuration is shown.pd data are from \cite{zejma97}}}}
\end{center}
\end{figure}

\begin{figure}[hbt]
\vskip 0.0 true cm  
\begin{center}
\centerline{\hskip 0.2 true cm \psfig{file=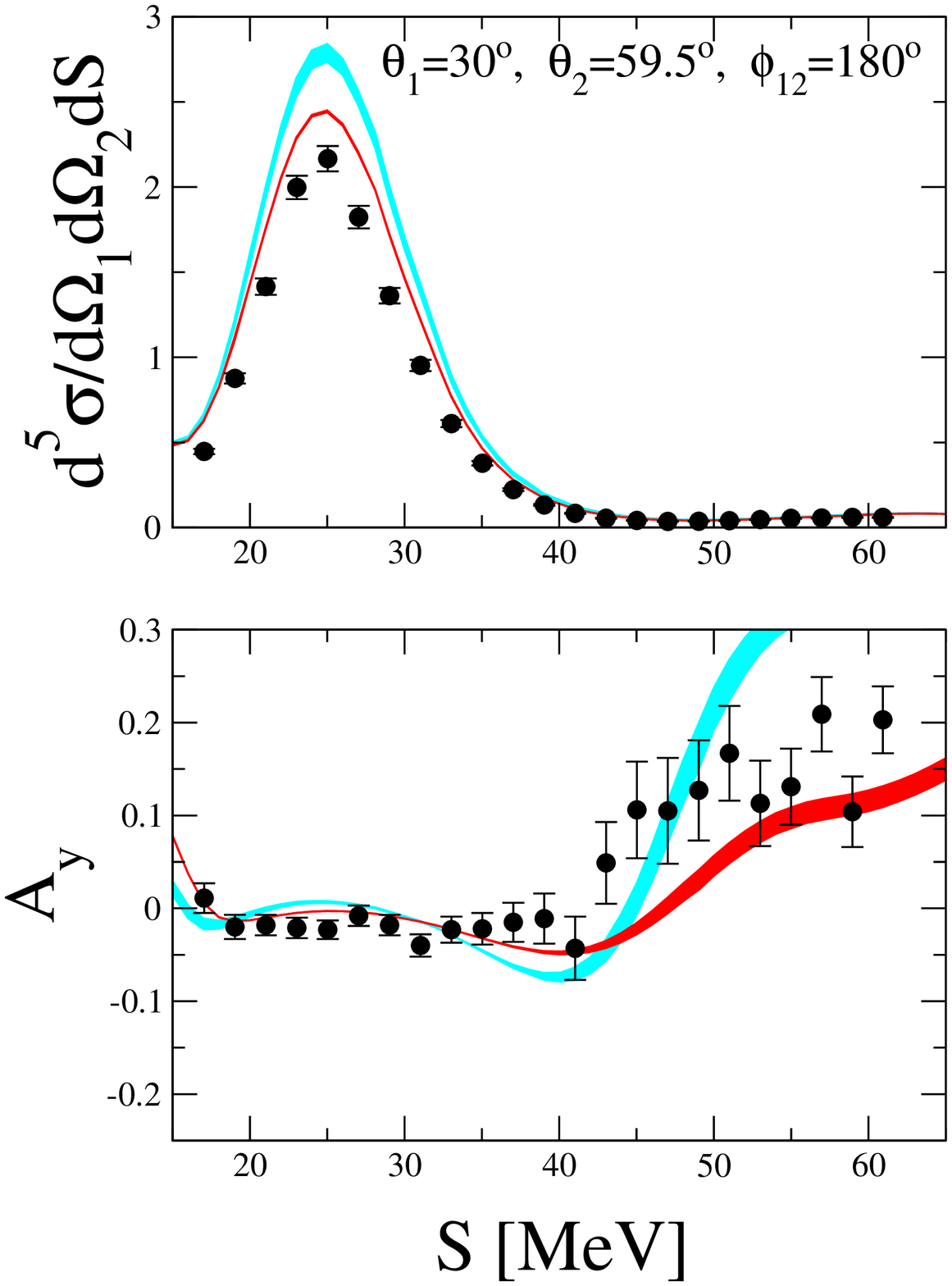,width=14cm}}
\centerline{\parbox{14cm}{
\caption{\label{fig10} $nd$ break up cross section in [mb MeV$^{-1}$ sr$^{-2}$] and nucleon analyzing power 
along the kinematical locus $S$ (in MeV) 
at 65 MeV in comparison to predictions at NLO (light shaded band) and NNLO (dark shaded band) in chiral effective
field theory. Quasi--free scattering configuration is shown. $pd$ data are from \cite{zejma97}}}}
\end{center}
\end{figure}

\begin{figure}[hbt]
\vskip 0.0 true cm  
\begin{center}
\centerline{\hskip 0.2 true cm \psfig{file=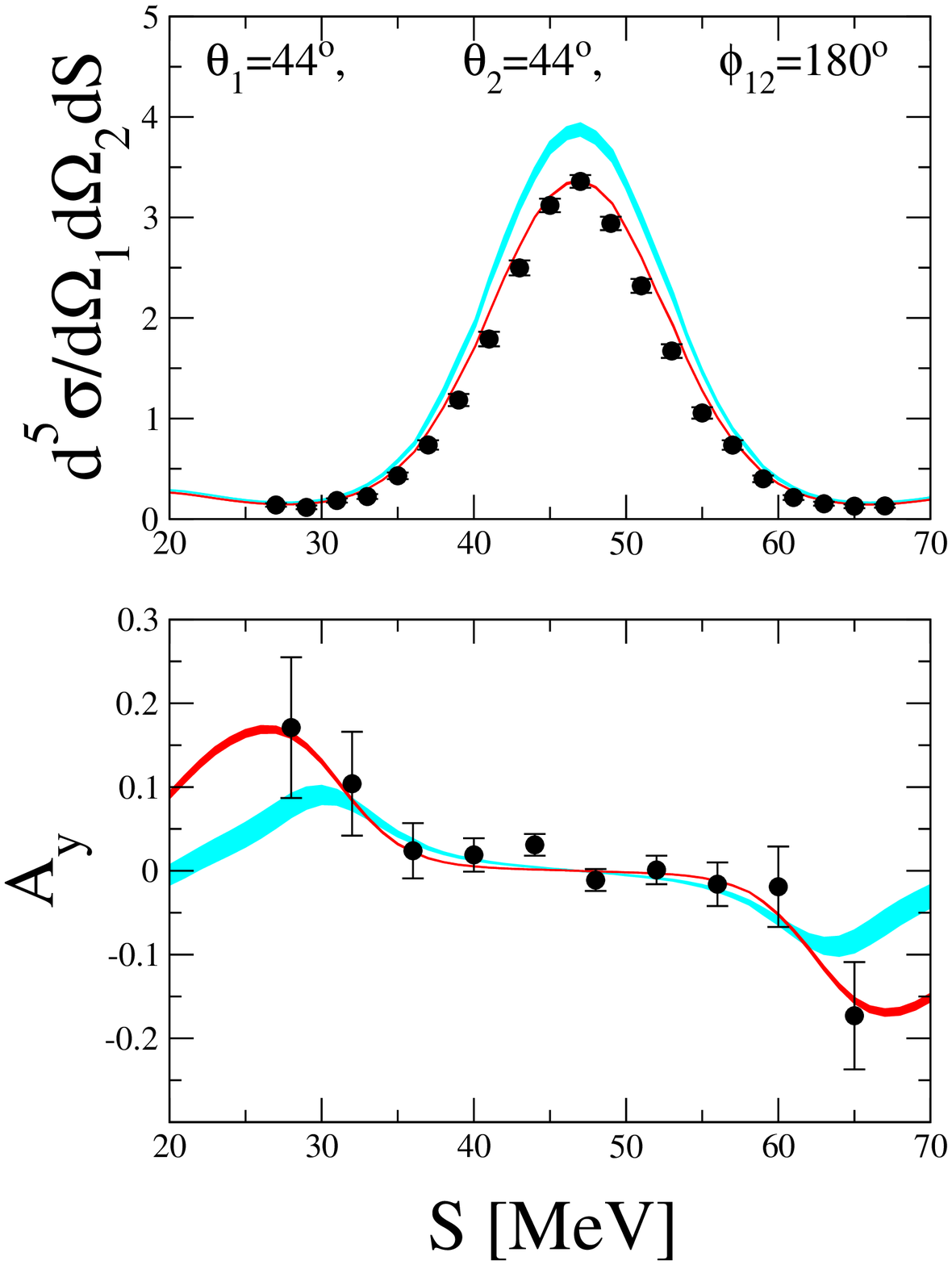,width=14cm}}
\centerline{\parbox{14cm}{
\caption{\label{fig11} $nd$ break up cross section in [mb MeV$^{-1}$ sr$^{-2}$] and nucleon analyzing power 
along the kinematical locus $S$ (in MeV) 
at 65 MeV in comparison to predictions at NLO (light shaded band) and NNLO (dark shaded band) in chiral effective
field theory. Quasi--free scattering configuration is shown. $pd$ data are from \cite{zejma97}}}}
\end{center}
\end{figure}

\begin{figure}[hbt]
\vskip 0.0 true cm  
\begin{center}
\centerline{\hskip 0.2 true cm \psfig{file=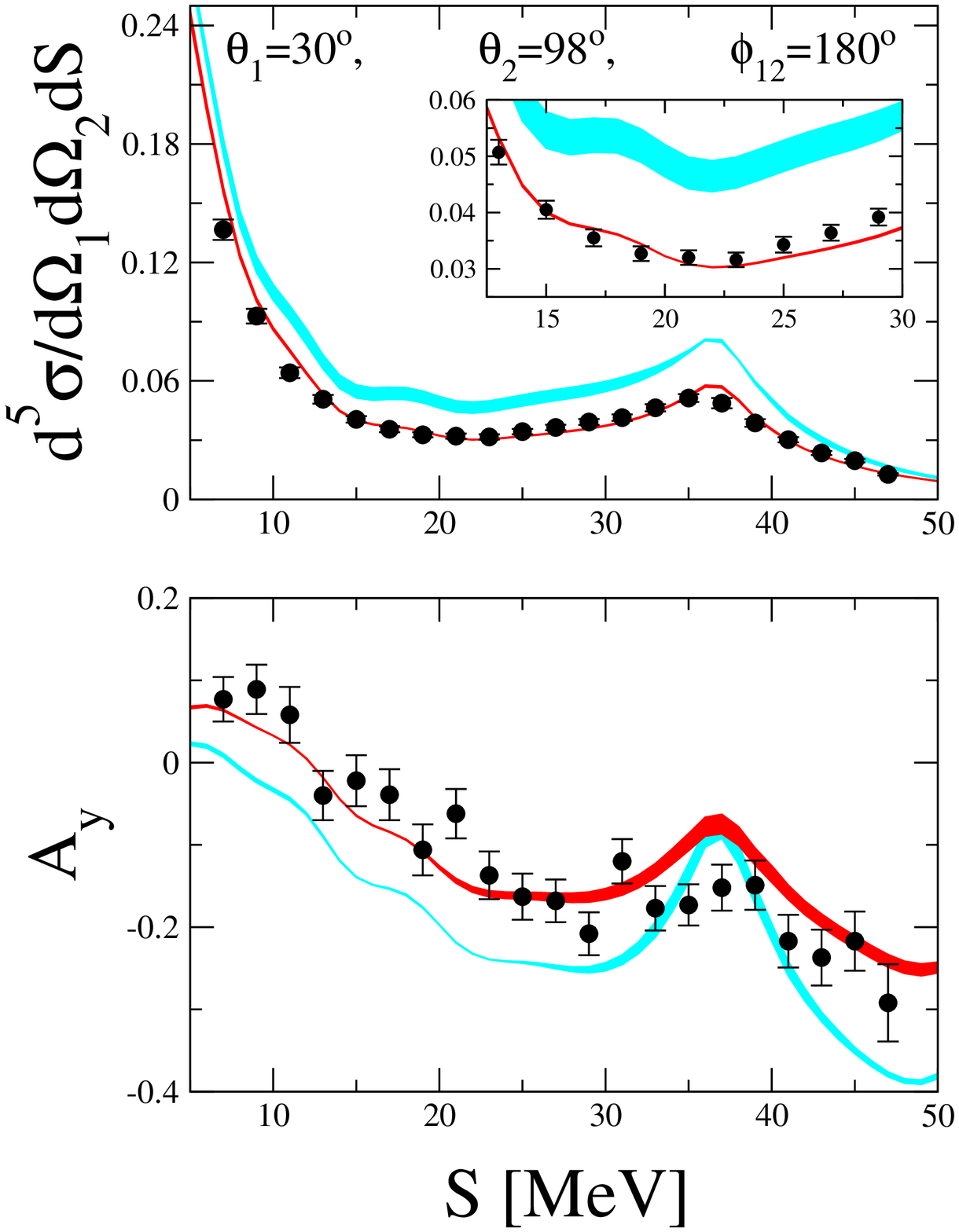,width=14cm}}
\centerline{\parbox{14cm}{
\caption{\label{fig12} $nd$ break up cross section in [mb MeV$^{-1}$ sr$^{-2}$] and nucleon analyzing power 
along the kinematical locus $S$ (in MeV) 
at 65 MeV in comparison to predictions at NLO (light shaded band) and NNLO (dark shaded band) in chiral effective
field theory. Collinear configuration is shown. $pd$ data are from \cite{allet94}}}}
\end{center}
\end{figure}

\begin{figure}[hbt]
\vskip 0.0 true cm  
\begin{center}
\centerline{\hskip 0.2 true cm \psfig{file=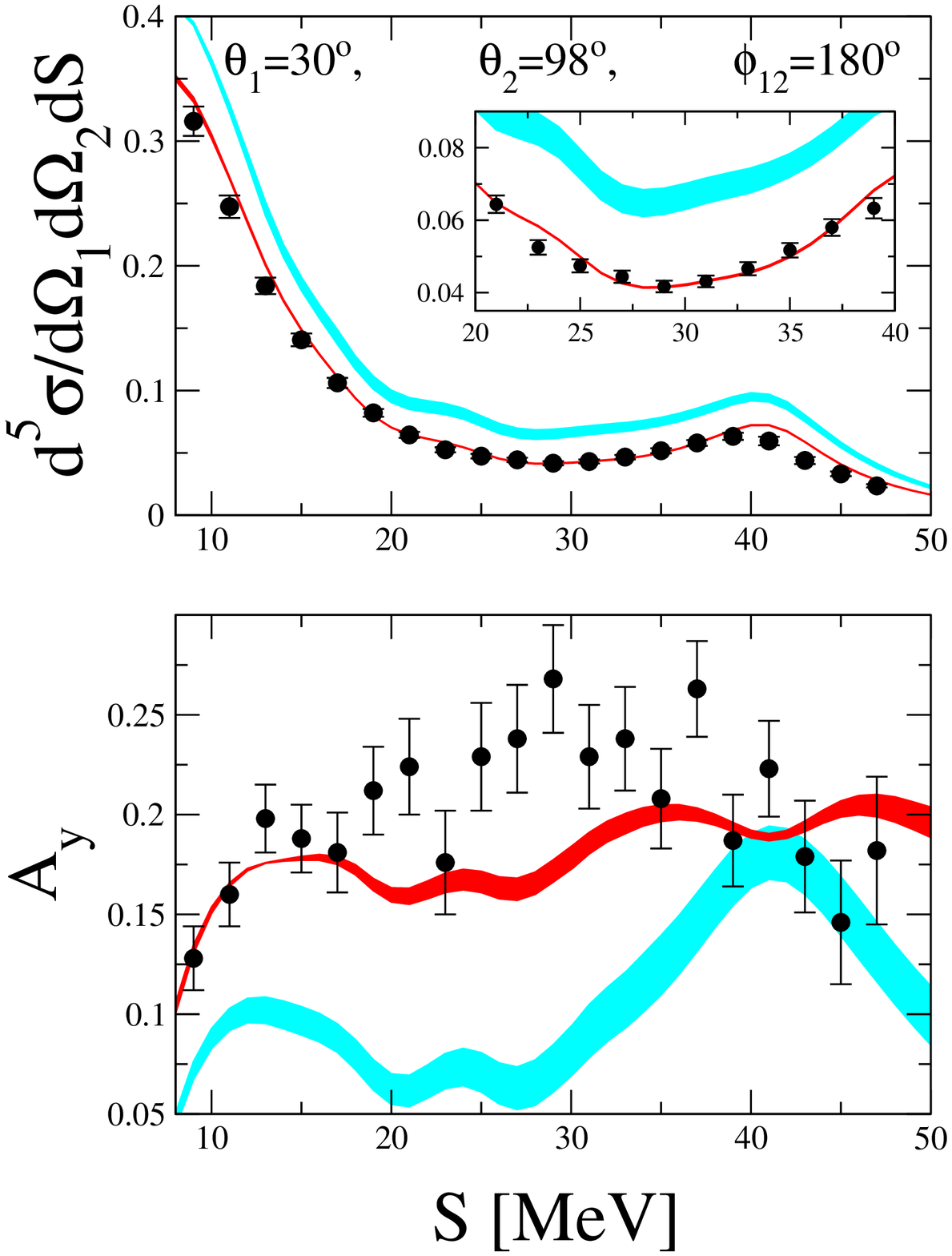,width=14cm}}
\centerline{\parbox{14cm}{
\caption{\label{fig13} $nd$ break up cross section in [mb MeV$^{-1}$ sr$^{-2}$] and nucleon analyzing power 
along the kinematical locus $S$ (in MeV) 
at 65 MeV in comparison to predictions at NLO (light shaded band) and NNLO (dark shaded band) in chiral effective
field theory. Collinear configuration is shown. $pd$ data are from \cite{allet94}}}}
\end{center}
\end{figure}

\begin{figure}[hbt]
\vskip 0.0 true cm  
\begin{center}
\centerline{\hskip 0.2 true cm \psfig{file=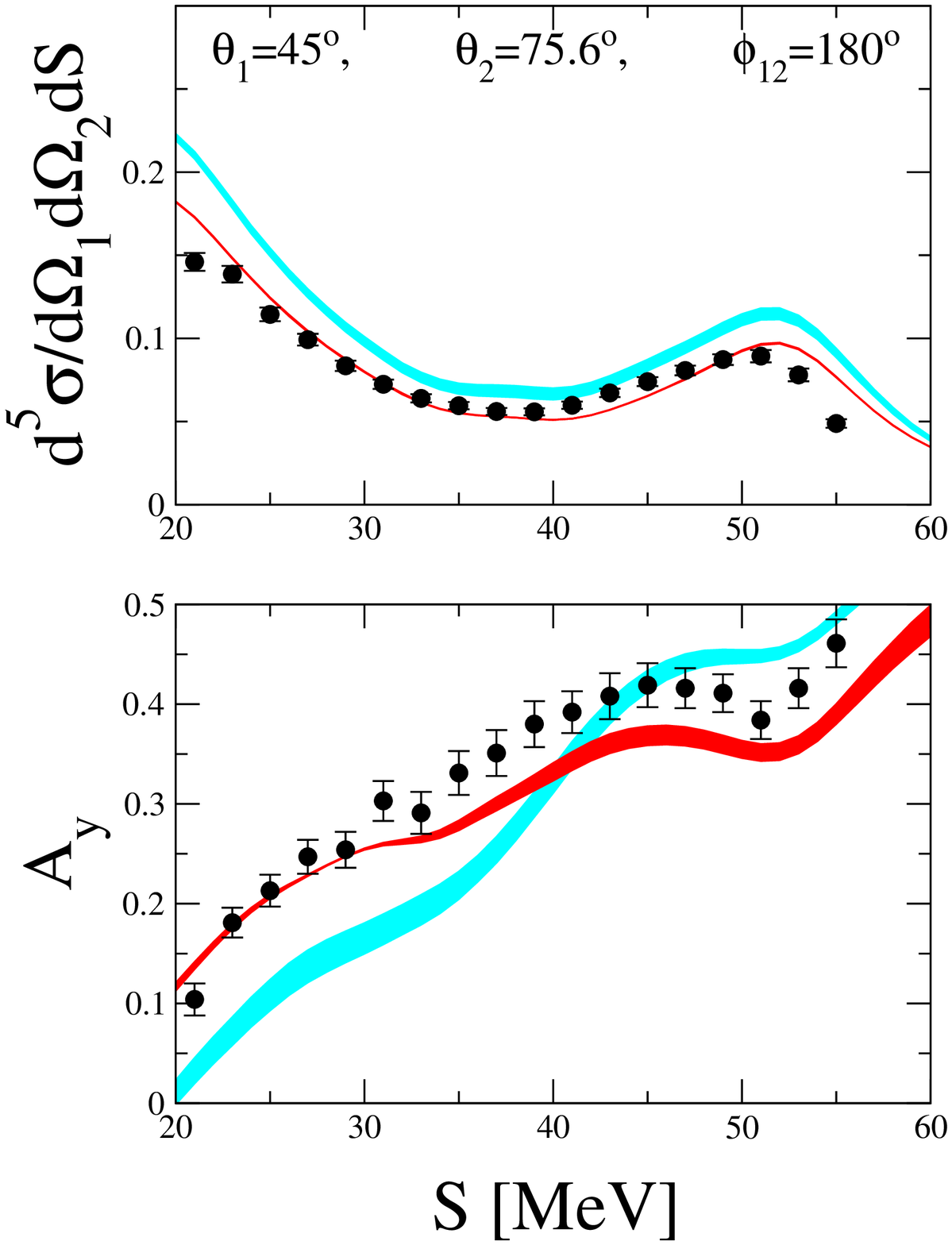,width=14cm}}
\centerline{\parbox{14cm}{
\caption{\label{fig14} $nd$ break up cross section in [mb MeV$^{-1}$ sr$^{-2}$] and nucleon analyzing power 
along the kinematical locus $S$ (in MeV) 
at 65 MeV in comparison to predictions at NLO (light shaded band) and NNLO (dark shaded band) in chiral effective
field theory. Collinear configuration is shown. $pd$ data are from \cite{allet94}}}}
\end{center}
\end{figure}

\begin{figure}[hbt]
\vskip 0.0 true cm  
\begin{center}
\centerline{\hskip 0.2 true cm \psfig{file=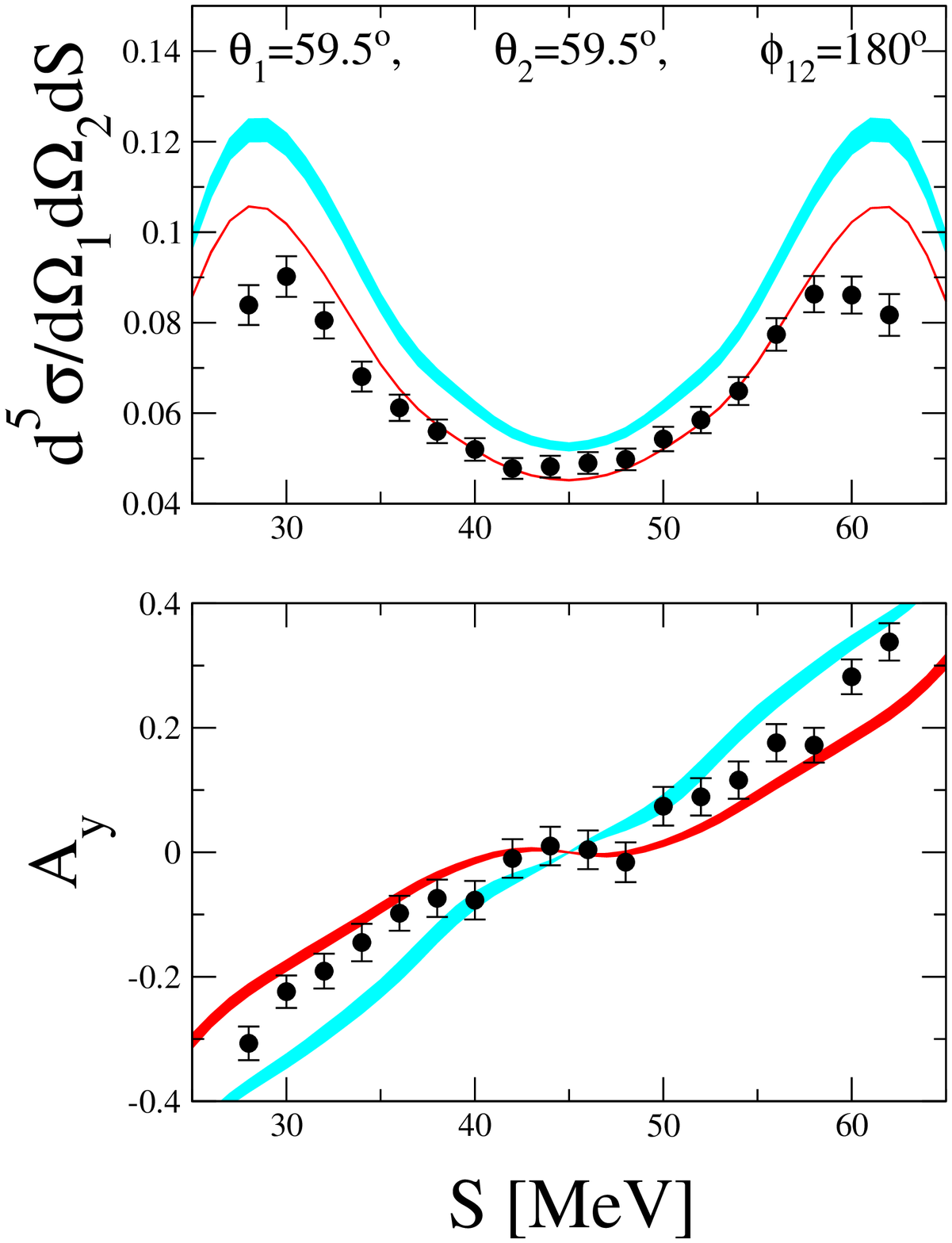,width=14cm}}
\centerline{\parbox{14cm}{
\caption{\label{fig15} $nd$ break up cross section in [mb MeV$^{-1}$ sr$^{-2}$] and nucleon analyzing power 
along the kinematical locus $S$ (in MeV) 
at 65 MeV in comparison to predictions at NLO (light shaded band) and NNLO (dark shaded band) in chiral effective
field theory. Collinear configuration is shown. $pd$ data are from \cite{allet94}}}}
\end{center}
\end{figure}

\begin{figure}[hbt]
\vskip 0.0 true cm  
\begin{center}
\centerline{\hskip 0.2 true cm \psfig{file=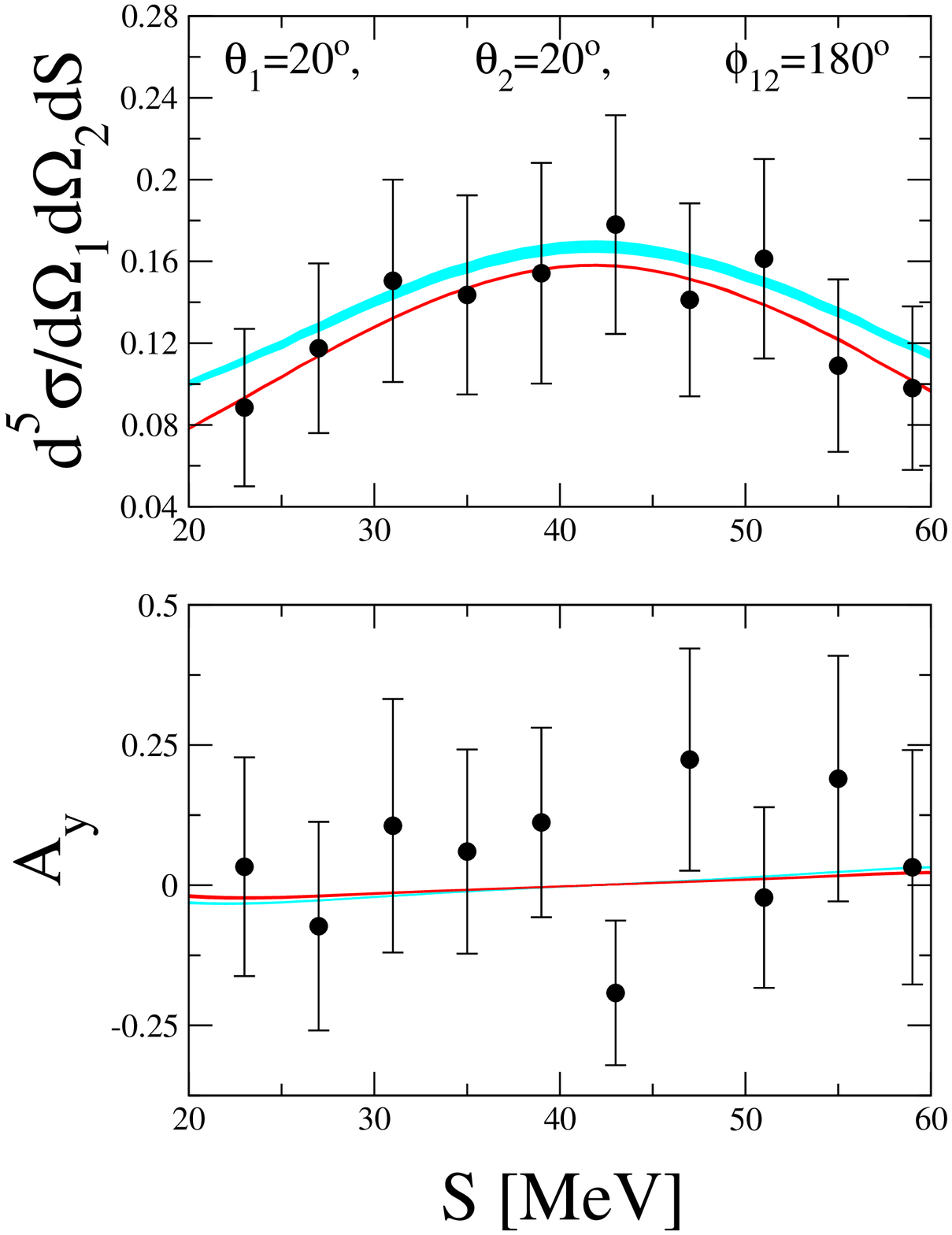,width=14cm}}
\centerline{\parbox{14cm}{
\caption{\label{fig16} $nd$ break up cross section in [mb MeV$^{-1}$ sr$^{-2}$] and nucleon analyzing power 
along the kinematical locus $S$ (in MeV) 
at 65 MeV in comparison to predictions at NLO (light shaded band) and NNLO (dark shaded band) in chiral effective
field theory. Unspecific configuration is shown. $pd$ data are from \cite{bodek01} }}}
\end{center}
\end{figure}

\begin{figure}[hbt]
\vskip 0.0 true cm  
\begin{center}
\centerline{\hskip 0.2 true cm \psfig{file=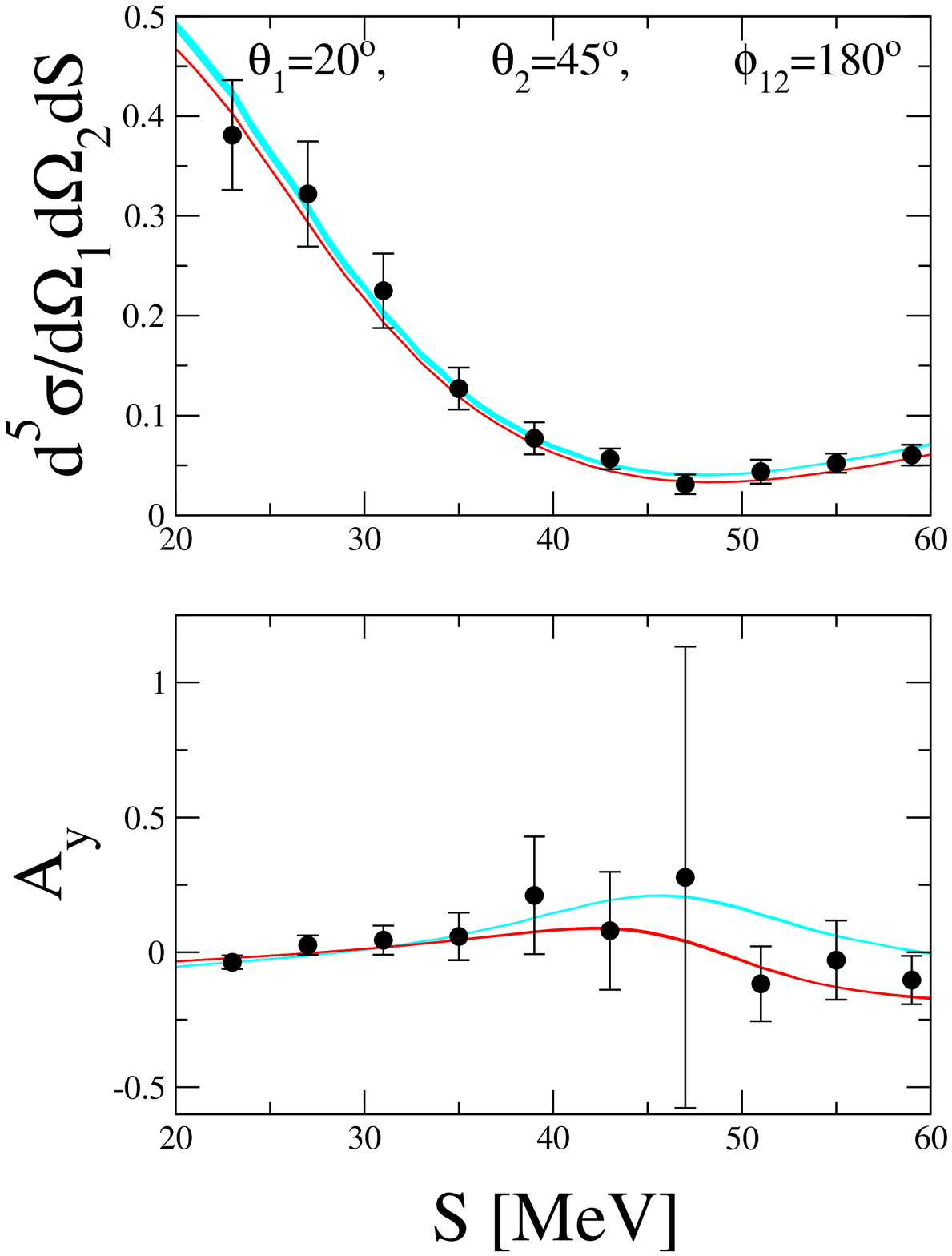,width=14cm}}
\centerline{\parbox{14cm}{
\caption{\label{fig17} $nd$ break up cross section in [mb
  MeV$^{-1}$ sr$^{-2}$] and nucleon analyzing power  
along the kinematical locus $S$ (in MeV) 
at 65 MeV in comparison to predictions at NLO (light shaded band) and
NNLO (dark shaded band) in chiral effective 
field theory. Unspecific configuration is shown. $pd$ data are from \cite{bodek01}  }}}
\end{center}
\end{figure}

\begin{figure}[hbt]
\vskip 0.0 true cm  
\begin{center}
\centerline{\hskip 0.2 true cm \psfig{file=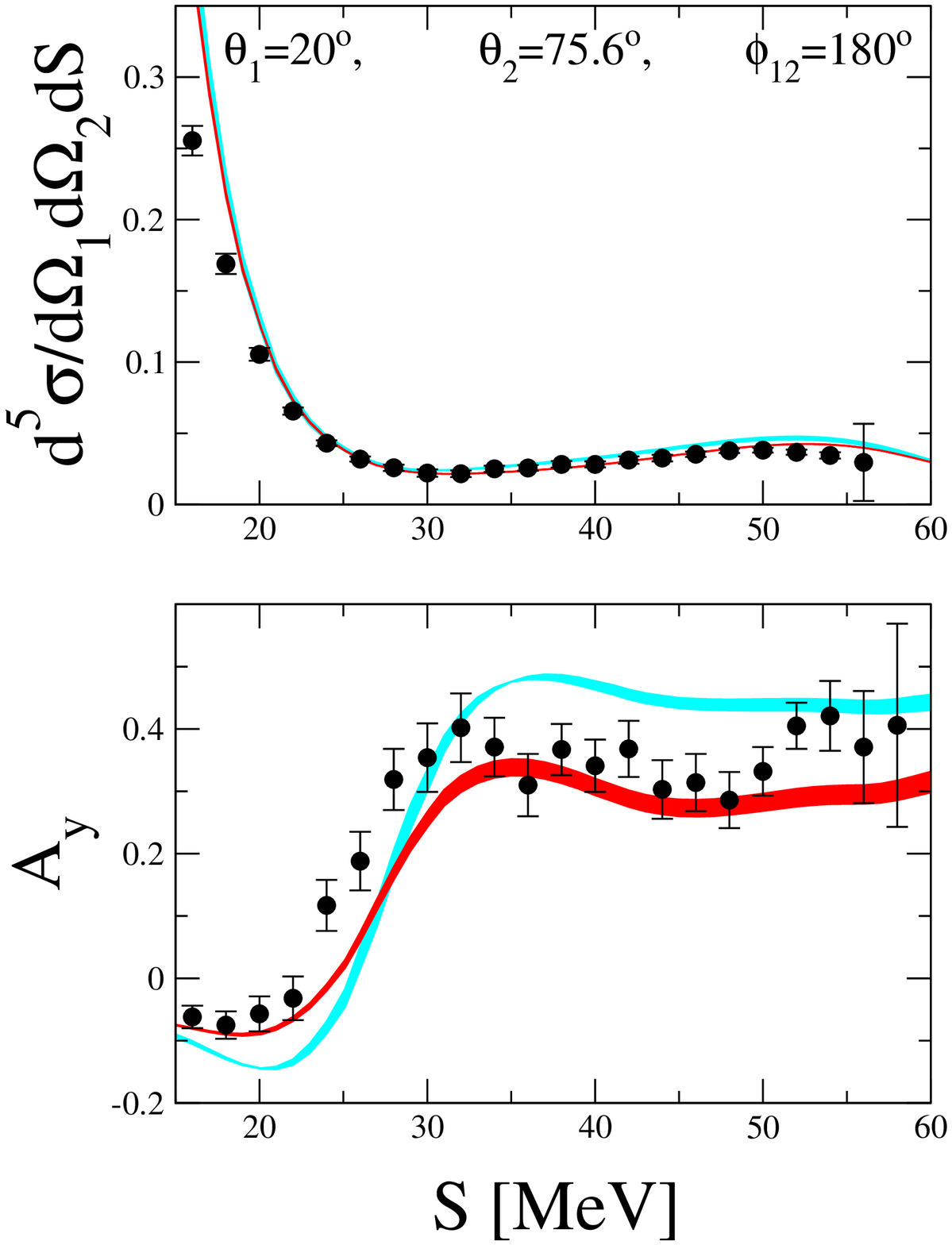,width=14cm}}
\centerline{\parbox{14cm}{
\caption{\label{fig18} $nd$ break up cross section  in [mb
  MeV$^{-1}$ sr$^{-2}$] and nucleon analyzing power  
along the kinematical locus $S$ (in MeV) 
at 65 MeV in comparison to predictions at NLO (light shaded band) and
NNLO (dark shaded band) in chiral effective 
field theory. Unspecific configuration is shown. $pd$ data are from \cite{bodek01} }}}
\end{center}
\end{figure}

\begin{figure}[hbt]
\vskip 0.0 true cm  
\begin{center}
\centerline{\hskip 0.2 true cm \psfig{file=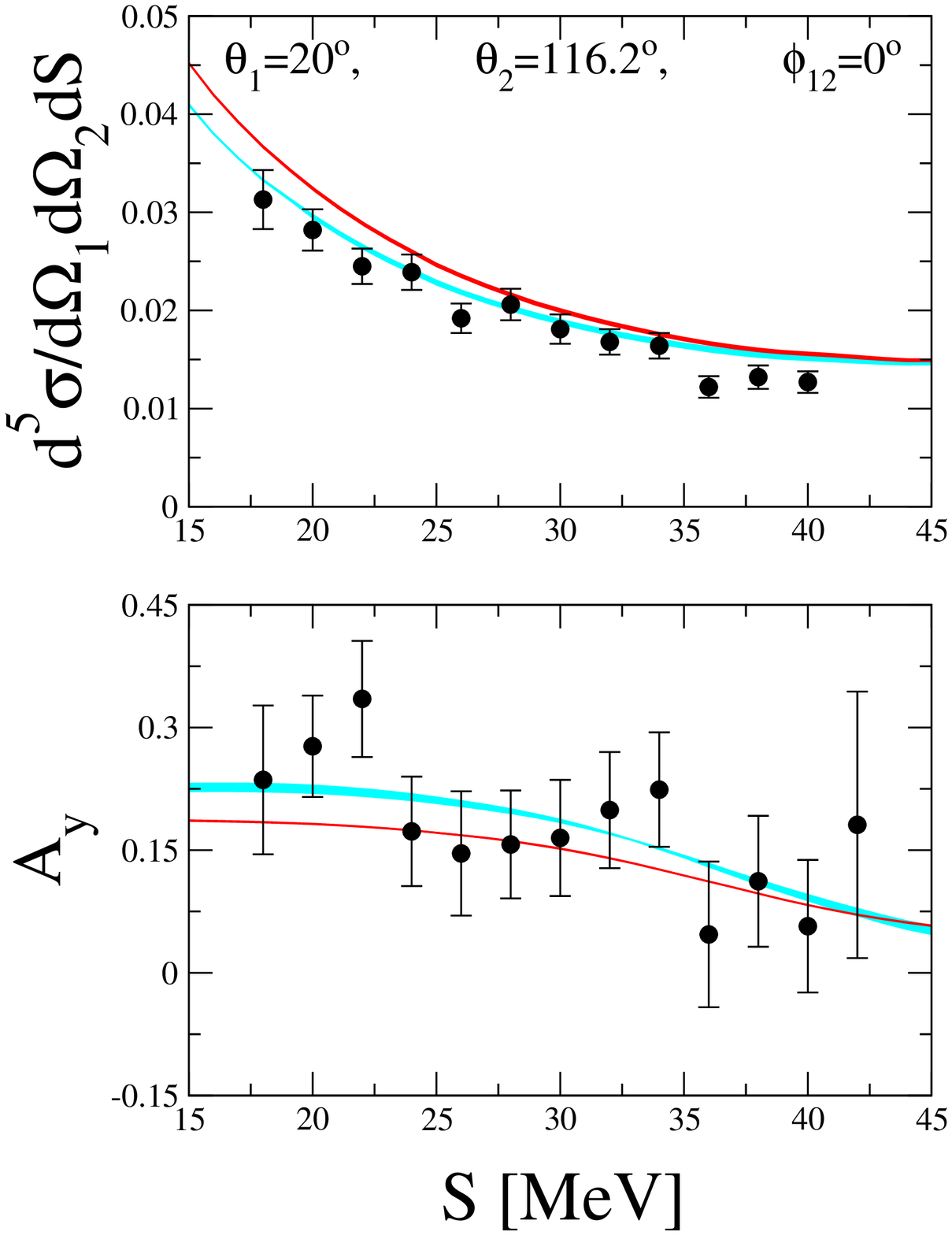,width=14cm}}
\centerline{\parbox{14cm}{
\caption{\label{fig19} $nd$ break up cross section in [mb
  MeV$^{-1}$ sr$^{-2}$] and nucleon analyzing power  
along the kinematical locus $S$ (in MeV) 
at 65 MeV in comparison to predictions at NLO (light shaded band) and
NNLO (dark shaded band) in chiral effective 
field theory. Unspecific configuration is shown. $pd$ data are from \cite{bodek01} }}}
\end{center}
\end{figure}

At 65~MeV we decided to present the results for the same configurations
as the ones studied recently in the context of phenomenological nuclear forces
\cite{kuros02a,kuros02b} in order to enable a direct comparison between these two 
different approaches.  We follow the lines of 
\cite{kuros02a,kuros02b} and include, in addition to the cross section, 
also $A_y$.
The situation at 65 MeV seems in general to be very promising as
documented in Figs.~\ref{fig8}-\ref{fig19}.  
The improvements in the description of the
data in going from NLO to NNLO are quite impressive.
It is interesting  that sometimes in case of  $A_y$ 
the band width at NNLO is still relatively large, which indicates
that this observable might get significant corrections at higher orders.
In Figs.~\ref{fig10} and \ref{fig15}   
we fail to describe the cross section in part of the  $S$-range. 
The reason is not known to us. These observables 
also change visibly, when going from NLO to NNLO. Here we cannot claim 
that convergence with respect to the chiral expansion is reached at NNLO.
For one of the configurations (see Fig.~\ref{fig13}), 
the step for $A_y$ going from NLO to NNLO is
dramatic and better data would be very welcome. 
Finally we point to two more cases in Figs.\ref{fig12} and \ref{fig13},
where the band width in the cross section shrinks nicely going
to NNLO and where the agreement with the  data is quite good.

In view of the quite good description of the 
Nd elastic and break up data at 65~MeV at NNLO 
and  of the good description of the NN data
up to 200~MeV, we 
are optimistic and expect to be able to 
describe the data at NNLO 
in the energy regime towards 100~MeV. 
From investigations based on phenomenological 
interactions \cite{witala01a,kuros02a}, we expect that there 3NF effects
become clearly visible at these higher energies. 
In addition, observables at these energies will probably be more 
sensitive to the structure of the 3N interaction. For example, the sensitivity of 
the cross section minimum to the value of $c_D$ observed at 65~MeV 
is expected to be magnified at higher energies. 
Therefore we call for more 
data at intermediate energies, which could be compared 
to predictions of chiral EFT. Notice also that the existence 
of such high--quality data will be of a crucial importance for
higher order calculations, where more parameters in the 3NF will 
have to be fixed from the data and a better accuracy in the theory will be reached.

\section{Summary and Conclusion}
\def\theequation{\arabic{section}.\arabic{equation}}

In summary we applied for the first time the complete
chiral EFT interaction at NNLO to the 3N and 4N bound states 
and to 3N scattering. 
We reexamined the 3NF of the chiral 
interaction at NNLO and used antisymmetrization 
to eliminate all parameters except two. 
We showed that these two parameters can be determined 
from the $^3$H BE and the $^2a_{nd}$ scattering length. 
For the time being the accuracy of the scattering 
length is not sufficient to perform a precise determination of 
these two parameters. However, the favorable description of 
$nd$ scattering data indicates that the values chosen in this work 
are in a reasonable range. 

We showed that the obtained parameter free Hamiltonian 
leads to a good description of the $\alpha$--particle 
BE. The theory thus seems to be applicable to this densely bound
system. It will be interesting to apply the 3N Hamiltonian 
to light nuclei, e.g. using the no-core shell model
technique \cite{navratil01,navratil02}.
 
Overall we observe a good description 
of the data at NNLO. Most of the low energy 
elastic scattering data (at 3 and 10~MeV) 
are described at both orders NLO and NNLO
showing convergence of the chiral expansion 
and agreement with the data.  
$A_y$ turns out to be a problematic observable as 
there is still no agreement with the data and the predictions 
for NLO and NNLO disagree. Whether this will be cured by $1/m$ 
corrections has to be studied in a forthcoming paper. 

At 65~MeV the situation is also very promising. In general,
we observe that the NNLO predictions  move 
towards or onto the data, while the NLO results deviate significantly 
from the data. 
In Ref.~\cite{epelbaum02a} we 
found that the NNLO interaction can describe 
the NN phase shifts up to energies of 200~MeV  neutron lab energy. 
Here we see that the extension of the energy range going to NNLO for the 
two-body system is continued in the few-body systems.
 
This study was based on the systematic expansion of the 
nuclear force according to chiral perturbation theory applied to the NN potential. 
We emphasize that the favorable agreement with the data, 
the stability of our predictions when going from NLO to NNLO, observed in 
most cases,  as well as 
the decreased cut--off dependence of the NNLO results indicate 
consistency of our calculations.
New $nd$ data in the energy range between 65~MeV and 100~MeV 
are highly welcome and would allow to draw quantitative 
conclusions on the range of validity of the NNLO approximation
as well as to probe the spin structure of the leading 3NF.
Such data would also be of a crucial importance for extending
the analysis to higher orders. 

In the next steps, we have to take into account the  
isospin breaking of the nuclear force.
Together with upcoming new data for the doublet $nd$ scattering 
length a much more accurate determination of the 
3NF parameters will than be possible.

\section*{Acknowledgments}
\def\theequation{\arabic{section}.\arabic{equation}}

We are very thankful to Alejandro Kievsky for supplying 
the $pd$ scattering observables based on AV18+Urbana~IX
and to Jacek Golak for checking the partial wave decomposition
of the 3NF and for the help in numerical problems.  
E.E. and A.N. would like to thank for the 
hospitality of the Science Center in Benasque, Spain,  
where a large part of this paper has been written. 
This work has been partially supported by the Deutsche 
Forschungsgemeinschaft (E.E.), the U.S. National 
Science Foundation under Grant \#PHY-0070858 (A.N.) and 
the Polish Committee for Scientific Research under Grant \#2P03B02818 (H.W.).
The numerical calculations have been performed on the Cray T3E 
and Cray SV1 of the NIC, J\"ulich, Germany.

\renewcommand{\theequation}{A-\arabic{equation}}
\setcounter{equation}{0}  
\section*{{\bf Appendix A.} Partial wave decomposition of the chiral 3NF}  

Here, we give the explicit formula for the partial wave decomposition of the 
chiral 3NF. Since the partial wave decomposition of the 
TPE 3NF is already discussed e.g.~in \cite{huber97b}, we only 
concentrate here on the remaining contributions to the NNLO 3NF due to the 
OPE and contact term in eq.~(\ref{3nfrest}). For general details on the 
partial wave decomposition in the 3N system the reader is addressed to Ref.~\cite{glocklefb}.
As already pointed out before,
we usually decompose 3NFs into three parts according to eq.~(\ref{decomp}). 
In the following we give expressions for one such 
part $V_{\rm 3NF}^{(i)}$. For the OPE term we find:
\beqa
\label{pwdOPE}
_i \langle p q \alpha | V_{\rm 3NF, \, \, OPE}^{(i)} | p ' q ' \alpha ' \rangle_i 
&=& - \frac{9 D g_A}{4 f_\pi^2} (4 \pi)^2 \delta_{JJ'} \, \delta_{MM'} \, \delta_{TT'} \delta_{M_T M_T '}
\delta_{l0} \delta_{l'0} \delta_{sj} \delta_{s'j'} \left[ 1 + (-1)^{s + s' + t + t'} \right] \nn
&&\times \sqrt{\hat s ' \hat j \hat I \hat I ' \hat t \hat t '} \, (-1)^{j+J+s-I+T+\frac{1}{2}} \, 
\left\{ \begin{array}{ccc} \frac{1}{2} & t & T \\ t ' & \frac{1}{2} & 1 \end{array} \right\} 
\, \left\{ \begin{array}{ccc} 1 & \frac{1}{2} & \frac{1}{2} \\ \frac{1}{2} & t & t ' \end{array} \right\} \nn
&& \times 
\left\{ \begin{array}{ccc} I & j & J \\ j' & I' & 1 \end{array} \right\} 
\, \left\{ \begin{array}{ccc} 1 & \frac{1}{2} & \frac{1}{2} \\ \frac{1}{2} & s & s ' \end{array} \right\}
\, \sum_{k_1 = 0,2} \, \sqrt{\hat k_1} \, (1\, 1\, k_1, \, \, 0\, 0\, 0 ) \\
&& \times 
\sqrt{(2 k_1 + 1) \mbox{!}}\,
\left\{ \begin{array}{ccc} 1 & k_1 & 1 \\ \frac{1}{2} & \lambda '  & I ' \\
\frac{1}{2} & \lambda & I \end{array} \right\}  \sum_{l_1 + l_2 = k_1} \, {q '}^{l_1} \, {q }^{l_2}
\frac{1}{\sqrt{(2 l_1 )\mbox{!} \, (2 l_2)\mbox{!}}} \nn
&& \times \sum_k \, \hat k \, g_{k k_1} \,
\left\{ \begin{array}{ccc} \lambda & \lambda '  & k_1 \\ l_1 & l_2  & k \end{array} \right\} \,
(k\, l_1\, \lambda ', \, \, 0\, 0\, 0 )\, (k\, l_2\, \lambda, \, \, 0\, 0\, 0 )\,,
\nonumber
\eeqa
where 
\beq
g_{k k_1} = \int_{-1}^1 \, dx \, P_k (x) \frac{Q^2}{Q^{k_1} (Q^2 + M_\pi^2)}\,.
\eeq
Here $Q\equiv \sqrt{q^2 + {q '}^2 - 2 q q ' x}$ and $P_k (x)$ is a Legendre polynomial. 
Further, $\vec p$ and $\vec q$ ($\vec p \, '$ and  $\vec q \, '$)
are relative initial (final) Jacobi momenta in the pair $jk$, $j,k \neq i$,  and of the nucleon $i$ with respect to the 
pair $jk$, respectively. $l$ and $\lambda$ ($l '$ and $\lambda '$) denote the initial (final) relative 
orbital angular momenta within the pair $jk$, $j,k \neq i$,  and between the nucleon $i$ with respect to the 
pair $jk$. The initial (final) spin of the subsystem $jk$,  $j,k \neq i$, is denoted by $s$ ($s '$). In addition, 
$l$ and $s$ ($l '$ and $s '$) are coupled to the total subsystem angular momentum $j$ ($j'$), and $\lambda$
($\lambda '$) and $s_i=\frac{1}{2}$ to the total spectator angular momentum $I$ ($I '$), which finally combine 
to $J$ ($J '$) accompanied by $M$ ($M '$).  The total isospin quantum numbers $T \, M_T$ ($T' \, M_T '$) 
are constructed analogously:
$ | ( t \frac{1}{2} ) T M_T \rangle$ ($ | ( t ' \frac{1}{2} ) T ' M_T '\rangle$).
We also introduced a convenient abbreviation
\beq
\hat l \equiv 2 l +1\,.
\eeq
For the contact term in the second line of eq.~(\ref{3nfrest}) we find:
\beqa
\label{pwdcont}
_i \langle p q \alpha | V_{\rm 3NF, \, \, cont}^{(i)} | p ' q ' \alpha ' \rangle_i 
&=&  6 E  (4 \pi)^2 \delta_{JJ'} \, \delta_{MM'} \, \delta_{TT'} \delta_{M_T M_T '}
\delta_{l0} \delta_{\lambda 0} \delta_{l'0} \delta_{\lambda ' 0} \delta_{sj} \delta_{s'j'} 
\delta_{I \frac{1}{2}} \delta_{I ' \frac{1}{2}} \delta_{t t'} \delta_{ss'} \nn
&& \times  (-1)^{t + 1} \, \left\{ \begin{array}{ccc} \frac{1}{2} & \frac{1}{2} & t \\ \frac{1}{2} & \frac{1}{2} 
& 1 \end{array} \right\}  \,.
\eeqa

\vfill\eject
  
\bibliography{literatur}

\begin{thebibliography}{10}

\bibitem{weinberg90}
S.~Weinberg.
\newblock {\em Phys. Lett.}, B251:288, 1990.

\bibitem{weinberg91}
S.~Weinberg.
\newblock {\em Nucl. Phys.}, B363:3, 1991.

\bibitem{ordonez96}
{C. Ord\'{o}\~{n}ez}, {L. Ray }, and {U. van Kolck}.
\newblock {\em Phys. Rev. C}, 53:2086, 1996.

\bibitem{tamm45}
{I.~Tamm}.
\newblock {\em J. Phys., U.S.S.R.}, 9:449, 1945.

\bibitem{dancoff50}
{S.M.~Dancoff}.
\newblock {\em Phys. Rev.}, 78:382, 1950.

\bibitem{friar99b}
{J.L. Friar}.
\newblock {\em Phys.Rev. C}, 60:034002, 1999.

\bibitem{epelbaoum98b}
{{E. Epelbaoum}, {W. Gl\"ockle}, {Ulf-G. Mei{\ss}ner}}.
\newblock {\em Nucl. Phys.}, A637:107, 1998.

\bibitem{okubo54}
{S.~Okubo}.
\newblock {\em Progr. Theor. Phys., Japan}, 12:603, 1954.

\bibitem{epelbaum00}
{E. Epelbaum, W. Gl\"ockle, Ulf-G. Mei{\ss}ner}.
\newblock {\em Nucl. Phys.}, A671:295, 2000.

\bibitem{epelbaum02a}
{{E. Epelbaum}, {A. Nogga}, {W. Gl\"ockle}, {H. Kamada}, {Ulf.-G.
  Mei{\ss}ner},{H. Wita{\l}a}}, 2002.
\newblock nucl-th/0201064, accepted for publication in Eur. Phys. J. A.

\bibitem{stoks94}
{V.G.J. Stoks}, {R.A.M. Klomp}, {C.P.F. Terheggen}, and {J.J. de Swart}.
\newblock {\em Phys. Rev. C}, 49:2950, 1994.

\bibitem{kolck94}
{U. van Kolck}.
\newblock {\em Phys. Rev. C}, 49:2932, 1994.

\bibitem{yang86}
{S.N.~Yang, W. Gl\"ockle}.
\newblock {\em Phys. Rev. C}, 33:1774, 1986.

\bibitem{gasser79}
{J. Gasser} and {A. Zepeda}.
\newblock {\em Nucl. Phys.}, B174:445, 1980.

\bibitem{eden96}
{J.A. Eden} and {M.F. Gari}.
\newblock {\em Phys. Rev. C}, 53:1510, 1996.

\bibitem{epelbaum01a}
{E. Epelbaum, H. Kamada, A. Nogga, H. Wita{\l}a, W. Gl\"ockle, Ulf-G.
  Mei{\ss}ner}.
\newblock {\em Phys. Rev. Lett.}, 86:4787, 2001.

\bibitem{huber01}
{D. H\"uber, J.L.~Friar, A. Nogga, H. Wita{\l}a, U. van Kolck}.
\newblock {\em Few Body Systems}, 30:95, 2001.

\bibitem{bernard95}
{N. Kaiser} {V. Bernard} and {Ulf-G. Mei{\ss}ner}.
\newblock {\em Int. J. Mod. Phys.}, E4:193, 1995.

\bibitem{friar99a}
{J.L. Friar, H. H\"uber, U. van Kolck}.
\newblock {\em Phys.Rev. C}, 59:53, 1999.

\bibitem{bedaque00}
{P.F.~Bedaque, H.-W.~Hammer, U.~van~Kolck}.
\newblock {\em Nucl.Phys.}, A676:357, 2000.

\bibitem{friar97}
{J.L. Friar}.
\newblock {\em Few Body Systems}, 22:161, 1997.

\bibitem{epelbaum02b}
{{E. Epelbaum}, {Ulf.-G. Mei{\ss}ner}, {W. Gl\"ockle}, {Ch. Elster}}.
\newblock {\em Phys. Rev. C}, 65:044001, 2002.

\bibitem{glocklefb}
Walter Gl\"ockle.
\newblock {\em The Quantum Mechanical Few-Body Problem}.
\newblock Springer-Verlag, Berlin, 1983.

\bibitem{nogga97}
{A. Nogga}, {D. H\"uber}, {H. Kamada}, and {W. Gl\"ockle}.
\newblock {\em Phys. Lett.}, B409:19, 1997.

\bibitem{glockle96}
{W. Gl\"ockle, H. Wita\l a, D. H\"uber, H. Kamada, and J. Golak}.
\newblock {\em Phys. Rep.}, 274:107, 1996.

\bibitem{huber97a}
{D. H\"uber}, {H. Kamada}, {H. Wita\l a}, and {W. Gl\"ockle}.
\newblock {\em Acta Phys. Polonica}, B28:1677, 1997.

\bibitem{witala88}
{H. Wita\l a, Th. Cornelius, W. Gl\"ockle}.
\newblock {\em Few-Body Systems}, 3:123, 1988.

\bibitem{huber97b}
{D. H\"uber, H. Wita{\l}a, A. Nogga, W. Gl\"ockle, H. Kamada}.
\newblock {\em Few Body Systems}, 22:107, 1997.

\bibitem{hanhart00}
{C. Hanhart}, {U. van Kolck}, and {G.A. Miller}.
\newblock {\em Phys. Rev. Lett.}, 85:2905, 2000.

\bibitem{walzl01}
{Ulf-G. Mei{\ss}ner} {M. Walzl} and {E. Epelbaum}.
\newblock {\em Nucl. Phys.}, A693:663, 2001.

\bibitem{nogga02b}
{ {A. Nogga}, {H. Kamada}, {W. Gl\"ockle}, {B.R. Barrett}}.
\newblock {\em Phys. Rev. C}, 65:054003, 2002.

\bibitem{wiringa95}
{R.B. Wiringa}, {V.G.J. Stoks}, and {R. Schiavilla}.
\newblock {\em Phys. Rev. C}, 51:38, 1995.

\bibitem{pudliner97}
{B.S. Pudliner}, {V.R. Pandharipande}, {J. Carlson}, {Steven C. Pieper}, and
  {R.B. Wiringa}.
\newblock {\em Phys. Rev. C}, 56:1720, 1997.

\bibitem{machleidt01a}
{R. Machleidt}.
\newblock {\em Phys. Rev. C}, 63:024001, 2001.

\bibitem{coon01}
{S.A. Coon} and {H.K. Han}.
\newblock {\em Few Body Systems}, 30:131, 2001.

\bibitem{phillips68}
{A.C. Phillips}.
\newblock Phillips linie.
\newblock {\em Nucl.Phys.}, A107:209, 1968.

\bibitem{friar86b}
{J.L. Friar}.
\newblock {\em Few-Body Systems, Suppl.}, 1:94, 1986.

\bibitem{bedaque99}
{P.F.~Bedaque, H.-W.~Hammer, U.~van~Kolck}.
\newblock {\em Nucl.Phys.}, A646:444, 1999.

\bibitem{bedaque02}
{P.F.~Bedaque, U.~van~Kolck}.
\newblock {\em to appear in Ann. Rev. Nucl. Part. Sci.}, 52, 2002.

\bibitem{zimmerpriv}
{O. Zimmer}.
\newblock private communication.

\bibitem{snowpriv}
{W.M. Snow}.
\newblock private communication.

\bibitem{yakubovsky67}
O.A. Yakubovsky.
\newblock {\em Sov. J. Nucl. Phys.}, 5:937, 1967.

\bibitem{nogga00}
{ {A. Nogga}, {H. Kamada}, {W. Gl\"ockle}}.
\newblock {\em Phys. Rev. Lett.}, 85:944, 2000.

\bibitem{tjon75}
{J.A. Tjon}.
\newblock {\em Phys. Lett. B}, 56:217, 1975.

\bibitem{kievskypriv}
A.~Kievsky.
\newblock private communication.

\bibitem{kievsky99}
{A. Kievsky}.
\newblock {\em Phys. Rev. C}, 60:034001, 1999.

\bibitem{koike87}
{Y.~Koike, J.~Haidenbauer}.
\newblock {\em Nucl. Phys.}, A463:365c, 1987.

\bibitem{witala94}
{H. Wita\l a, D. H\"uber, W. Gl\"ockle}.
\newblock {\em Phys. Rev. C}, 49:R 14, 1994.

\bibitem{witala01a}
{{H. Wita{\l}a}, {W. Gl\"ockle}, {J. Golak}, {A. Nogga}, {H. Kamada}, {R.
  Skibi\'{n}ski}} and {J. Kuro\'{s}-\.{Z}o{\l}nierczuk}.
\newblock {\em Phys. Rev. C}, 63:024007, 2001.

\bibitem{epelbaum01b}
{{E. Epelbaum}, {H. Kamada}, {A. Nogga}, {H. Wita{\l}a}, {W. Gl\"ockle}, and
  {Ulf-G. Mei{\ss}ner}}.
\newblock {\em Nucl. Phys.}, A689:111, 2001.

\bibitem{kievsky98}
{A. Kievsky}, {M. Viviani}, {S. Rosati}, {D. H\"uber}, {W. Gl\"ockle},
  {H.~Kamada}, {H.~Wita{\l}a}, and {J. Golak}.
\newblock {\em Phys. Rev. C}, 58:3085, 1998.

\bibitem{tornow98}
{W. Tornow, H. Wita\l a, A. Kievsky}.
\newblock {\em Phys. Rev. C}, 57:555, 1998.

\bibitem{Kievsky01}
{A. Kievsky}, {M. Viviani}, and {S. Rosati}.
\newblock {\em Phys. Rev. C}, 64:024002, 2001.

\bibitem{kuros02a}
{{J. Kuro\'{s}-\.{Z}o{\l}nierczuk}, H. Wita{\l}a, J. Golak, H. Kamada, A.
  Nogga, {R. Skibi\'{n}ski}, W. Gl\"ockle }, 2002.

\bibitem{kuros02b}
{{J. Kuro\'{s}-\.{Z}o{\l}nierczuk}, H. Wita{\l}a, J. Golak, H. Kamada, A.
  Nogga, {R. Skibi\'{n}ski}, W. Gl\"ockle }, 2002.

\bibitem{howell98a}
{C.R. Howell et al.}
\newblock {\em Nucl. Phys.}, A631:692c, 1998.

\bibitem{zhou01}
{Z. Zhou et al.}
\newblock {\em Nucl. Phys.}, A684:545c, 2001.

\bibitem{tachikawa01}
{Y. Tachikawa}, {T. Yagita}, {S. Minami}, {T. Ishida}, {K. Tsuruta}, and {K.
  Sagara}.
\newblock {\em Nucl. Phys.}, A684:583c, 2001.

\bibitem{navratil01}
{P. Navr\'atil, J.P. Vary, W.E. Ormand, B.R. Barrett}.
\newblock {\em Phys. Rev. Lett.}, 87:172502, 2001.

\bibitem{navratil02}
{P. Navr\'atil, W.E. Ormand }.
\newblock {\em Phys. Rev. Lett.}, 88:152502, 2002.

\bibitem{shimizu95}
{S. Shimizu {\it et al. }}.
\newblock {\em Phys. Rev. C}, 52:1193, 1995.

\bibitem{sagara94}
{K. Sagara {\it et al. }}.
\newblock {\em Phys. Rev. C}, 50:576, 1994.

\bibitem{mcaninch93}
{J.E. McAninch {\it et al. }}.
\newblock {\em Phys. Lett.}, B307:13, 1993.

\bibitem{rauprich88}
{G. Rauprich} et~al.
\newblock {\em Few-Body Systems}, 5:67, 1988.

\bibitem{sperisen84}
{F. Sperisen} et~al.
\newblock {\em Nucl. Phys.}, A422:81, 1984.

\bibitem{howell87}
{C. R. Howell {\it et al.}}
\newblock {\em Few Body Systems}, 2:19, 1987.

\bibitem{witala93}
{H. Wita\l a} et~al.
\newblock {\em Few Body Systems}, 15:67, 1993.

\bibitem{rauprich91}
{G. Rauprich} et~al.
\newblock {\em Nucl. Phys.}, A535:313, 1991.

\bibitem{setze96}
{H. R. Setze} et~al.
\newblock {\em Phys. Lett.}, B388:229, 1996.

\bibitem{strate89}
{J. Strate} et~al.
\newblock {\em Nucl. Phys.}, A501:51, 1989.

\bibitem{zejma97}
{J. Zejma} et~al.
\newblock {\em Phys. Rev. C}, 55:42, 1997.

\bibitem{allet94}
{M. Allet} et~al.
\newblock {\em Phys. Rev. C}, 50:602, 1994.

\bibitem{bodek01}
{K. Bodek} et~al.
\newblock {\em Few Body Systems}, 30:65, 2001.

\end{thebibliography}
\bibliographystyle{unsrt}

\clearpage
\centerline {{\large \bf FIGURES}}

\vspace{1cm}

\clearpage 

\clearpage 

\clearpage 

\clearpage 

\clearpage 

\clearpage 


\clearpage 

\clearpage 

\clearpage

\clearpage

\clearpage

\clearpage

\clearpage

\clearpage

\clearpage

\clearpage

\clearpage

\clearpage

\clearpage

\end{document}